\documentstyle[10pt,psfig,float]{article}

\def\be{\begin{equation}}
\def\en{\end{equation}}

\begin{document}
\baselineskip = 24pt

\Large

\begin{center}
{\bf Non-axisymmetric relativistic Bondi-Hoyle accretion 
onto a Kerr black hole}
\end{center}

\normalsize

\vspace{0.5cm}

\begin{center}

Jos\'e A. Font$^{1}$,
Jos\'e M$^{\underline{\mbox{a}}}$ Ib\'a\~nez$^{2}$ and
Philippos Papadopoulos$^{1}$

\vspace{0.2cm}

$^{1}$
Max-Planck-Institut f\"ur Gravitationsphysik \\
Albert-Einstein-Institut \\
Schlaatzweg 1, 14473 Potsdam, Germany \\

$^{2}$
Departamento de Astronom\'{\i}a y Astrof\'{\i}sica\\
Universidad de Valencia, 46100 Burjassot (Valencia), Spain \\

e-mail: font@aei-potsdam.mpg.de \\
e-mail: ibanez@scry.daa.uv.es \\
e-mail: philip@aei-potsdam.mpg.de

\end{center}

\vspace{1cm}

\begin{abstract}

\noindent

In our program of studying numerically the so-called Bondi-Hoyle
accretion in the fully relativistic regime, we present here first
results concerning the evolution of matter accreting supersonically
onto a rotating (Kerr) black hole. These computations generalize
previous results where the non-rotating (Schwarzschild) case was
extensively considered. We parametrize our initial data by the
asymptotic conditions for the fluid and explore the dependence 
of the solution on the angular momentum of the black hole. Towards 
quantifying the robustness of our numerical results, we use two 
different geometrical {\em foliations} of the black hole spacetime, the
standard form of the Kerr metric in Boyer-Lindquist coordinates as
well as its Kerr-Schild form, which is free of coordinate
singularities at the black hole horizon.  We demonstrate some
important advantages of using such {\it horizon adapted coordinate
systems}.

Our numerical study indicates that regardless of the value of the
black hole spin the final accretion pattern is always stable, leading
to constant accretion rates of mass and momentum. The flow is
characterized by a strong tail shock, which, unlike the Schwarzschild
case, is increasingly wrapped around the central black hole as the
hole angular momentum increases.  The rotation induced asymmetry in
the pressure field implies that besides the well known drag, the black
hole will experience also a {\em lift} normal to the flow direction. This
situation exhibits some analogies with the Magnus effect of classical 
fluid dynamics.

\end{abstract}


\newpage

\section{Introduction}

In previous work we have extensively studied, numerically, the
relativistic extension of the so-called Bondi-Hoyle accretion (Hoyle
and Lyttleton 1939, Bondi and Hoyle 1944) onto a Schwarzschild black
hole (Font and Ib\'a\~nez 1998a,b, FI98a,b in the following). This
type of accretion, also known as wind or hydrodynamic accretion,
appears when a homogeneous flow of matter at infinity moves
non-radially towards a compact object (the accretor).  The matter flow
inside the {\it accretion radius}, after being decelerated by a
conical shock -- if, asymptotically, flowing supersonically -- is
ultimately captured by the central black hole. Most of the material is
dragged toward the hole at its rear part.

The standard astrophysical scenario motivating such studies involves
mass transfer and accretion in a close binary system that characterize
compact X-ray sources. In particular, and closely related to the wind
accretion process, one may consider the case in which the primary
star, typically a blue supergiant, lies inside its Roche lobe and
loses mass via a stellar wind. This wind impacts on the orbiting
compact star and a bow-shaped shock front forms around it by the
action of its gravitational field.

Analytic studies of wind accretion started with the pioneering
investigations of Hoyle and Lyttleton (1939) and Bondi and Hoyle
(1944). Three decades later the problem was first numerically
investigated by Hunt (1971). Since then, contributions of a large
number of authors (see, e.g., Ruffert 1994, Benensohn, Lamb and Taam
1997 and the list of references in those publications) extended the
simplified (but still {\it globally} correct) analytic models. This
helped develop a thorough understanding of the hydrodynamic accretion
scenario, in its fully three-dimensional character. These
investigations revealed the
formation of accretion disks and the appearance of non-trivial phenomena
such as shock waves or {\it flip-flop} instabilities. Clearly,
important progress in the field was only possible through detailed and
reliable numerical work.

Most of the existing numerical work has focused on {\it
  non-relativistic} accretors, i.e., non-compact stars. In those cases
it suffices to perform a numerical integration of the equations of
Newtonian hydrodynamics. If, however, the accretor is a neutron star
or a black hole it is clear that relativistic effects become
increasingly more important and, hence, must be included in the
physical model and in the corresponding numerical scheme.  Newtonian
hydrodynamics is a valid approximation far from the compact star but
it no longer holds when studying the flow evolution close to the inner
boundary of the domain (the surface of the star). In the case of a
black hole, this boundary is ultimately placed at the event horizon.
Near that region the problem is intrinsically relativistic or even
ultrarelativistic according to the velocities involved (approaching
the speed of light), and the gravitational accelerations significantly
deviate from the Newtonian values.

An accurate numerical modeling of the aforementioned scenarios
requires a {\em general relativistic} hydrodynamical description which
at the same time is capable of handling extremely relativistic flows.
The utility of such methodology would extend to further interesting
astrophysical situations, like stellar collapse or coalescing compact
binaries. Recently, Banyuls et al. (1997) have proposed a new framework in
which the general relativistic hydrodynamic equations 
are written in conservation form to exploit,
numerically, their hyperbolic character. Taking advantage of the
hyperbolicity of the equations has proven to lead to an accurate
description of relativistic flows, in particular ultra-relativistic
flows with large bulk Lorentz factors.

The detailed description of accretion flows in the near horizon region, in
particular for {\em rotating} black holes depends crucially on the
coordinate language with respect to which quantities are expressed.
Coordinates adapted to observers at infinity lead to metric
expressions with singular appearance at the horizon.  In those
coordinates one is forced to locate the inner boundary {\it outside}
the horizon, which introduces the extraneous question of what
constitutes a sound choice. More importantly, singular systems
introduce, unwarranted, extreme dynamical behavior.  A simple example
is the behavior of the coordinate fluid velocity near the horizon of a
Schwarzschild black hole. In Schwarzschild coordinates it approaches
the speed of light causing the Lorentz factor to diverge and,
ultimately, the numerical code to terminate.  Papadopoulos and Font
(1998) (PF98 in the following) have recently shown that coordinates
adapted to the horizon region, and hence regular there, can greatly
simplify the integration of the general relativistic hydrodynamic
equations near black holes.  With those coordinates the innermost
radial boundary can be placed {\it inside} the horizon, allowing for a
clean treatment of the entire physical domain. The application of
this concept to rotating black holes was briefly outlined in
Font, Ib\'a\~nez and Papadopoulos (1998, FIP98 in the following), where we
focused on the important advantages of this new approach to the
numerical study of accretion flows.

For our study of relativistic hydrodynamic accretion we integrate the 
equations in the fixed background of the Kerr spacetime. 
We neglect the self-gravity of
the fluid as well as non-adiabatic processes such as viscosity or
radiative transfer. Our different initial models are parametrized
according to the value of the Kerr angular momentum per unit mass. The
asymptotic conditions of the flow at {\it infinity} (in practice a
sufficiently far distance from the black hole location), need only to
be imposed on the fluid velocity and sound speed (or Mach number,
indistinctly). We start fixing these quantities as well as the adiabatic
exponent of the perfect fluid equation of state,
focusing on the implications of the rotation of the hole in the final
accretion pattern. Additionally, we also analyze the influence of
varying the adiabatic index of the fluid on the flow morphology and
accretion rates for a rapidly rotating black hole.

As an important simplifying assumption, our numerical study is
restricted to the equatorial plane of the black hole.  Hence, we adopt
the ``infinitesimally thin'' accretion disk setup.  This is only
motivated by simplicity considerations, before attempting
three-dimensional studies. Existing Newtonian simulations of wind
accretion in cylindrical and Cartesian coordinates using the same
setup can be found in Matsuda et al (1991) or Benensohn et al (1997).
Within this assumption we are using a restricted set of equations,
where the vertical structure of the flow is assumed not to depend on
the polar coordinate. This requires that in the {\em immediate
  neighborhood} of the equator, vertical (polar) pressure gradients,
velocities and gravity (tidal) terms vanish. Those conditions are,
however, strictly correct at the equator for flows that are reflection
symmetric there. In particular, our dimensional simplification still
captures the most demanding aspect of the Kerr background, which is
encoded in the large azimuthal shift vector near the horizon.

The present investigation extends our previous studies (FI98a,b) of
Bondi-Hoyle accretion flows to account for rotating black holes with
arbitrary spins. We perform computations using both the standard
Boyer-Lindquist (BL) form of the metric as well as the Kerr-Schild (KS)
form. We develop the procedure for comparing the two computations
once a stationary flow pattern has been achieved.  The computations
reported here constitute the first simulations of non-axisymmetric
{\it relativistic} Bondi-Hoyle accretion flows onto rotating black holes.

Related work in the literature has explored the {\em potential flow}
approximation (Abrahams and Shapiro 1990). Within this formulation, in
which the hydrodynamic equations transform into a scalar second order
differential equation for a potential, Abrahams and Shapiro (1990)
computed a number of stationary and axisymmetric flows past a hard
sphere moving through an asymptotically homogeneous medium. For a
general polytropic equation of state the potential equation is
non-linear and elliptic but for the particular case $p=\rho$ ($p$
being the pressure and $\rho$ the density) the equation is linear. In
this case they could solve it, analytically, for steady-state flows
around hard spheres in Kerr (and Schwarzschild) geometries.
Additionally, recent numerical studies of hydrodynamical flows in the
Kerr spacetime, in the context of accretion discs, can be found in
Yokosawa (1993) or Igumensshchev and Beloborodov (1997).

The organization of the paper is as follows: in next Section ($\S 2$)
we present the system of equations of general relativistic
hydrodynamics written as a hyperbolic system of conservation
laws. They are specialized for the equatorial plane of the Kerr
metric. We write down the line element and the hydrodynamic equations
in both, the standard BL coordinates and the proposed KS system.
Transformations of the fluid quantities between the two systems are
shown here. Pertinent technical details are moved to the
Appendix. In addition, all numerical issues related to the code,
boundary conditions and initial setup are also described in Section
$\S 2$. The results of the simulations are presented and analyzed in
Section $\S 3$.  Those include the description of the flow morphology
and dynamics, the computation of the accretion rates of mass, linear
momentum and angular momentum and the comparison of different hydrodynamic
quantities in the two coordinate systems we use. Additionally,
we briefly describe in Section $\S 3$ the analogy between hydrodynamical 
flows past a rotating black hole and the Magnus effect of classical 
fluid dynamics.  Finally, Section $\S 4$ summarizes the main
conclusions of this work and outlines future directions.

\section{Equations and numerical issues}

\subsection{The Kerr metric in various coordinate systems}

In BL $(t,r,\theta,\phi)$ coordinates, the Kerr line element,
$ds^2=g_{\mu\nu} dx^{\mu} dx^{\nu}$, reads
\begin{eqnarray}
ds^2 = - { {\Delta - a^2\sin^2\theta} \over {\varrho^2} } dt^2 -
       2a { {2Mr\sin^2\theta} \over {\varrho^2} } dt d\phi     +
       { {\varrho^2} \over {\Delta} } dr^2                     + &
       \varrho^2 d\theta^2 +
       { {\Sigma} \over {\varrho^2} }
       \sin^2\theta d\phi^2,
\label{blform}
\end{eqnarray}
\noindent
with the definitions:
\begin{eqnarray}
\Delta &\equiv& r^2 - 2Mr + a^2,
\\
\varrho^2 &\equiv& r^2 + a^2\cos^2\theta,
\\
\Sigma &\equiv& (r^2+a^2)^2 - a^2\Delta\sin^2\theta,
\end{eqnarray}
\noindent
where $M$ is the mass of the hole and $a$ the black hole angular
momentum per unit mass ($J/M$). Throughout the paper we are using
geometrized units ($G=c=1$). Greek (Latin) indices run from 0 to 3 (1
to 3).  

This metric describes the spacetime exterior to a rotating and
non-charged black hole. It is characterized by the presence of an
azimuthal shift term, $\beta_{\phi}\equiv g_{t\phi}$. The
metric~(\ref{blform}) is singular at the roots of the equation
$\Delta=0$, which correspond to the horizons of a rotating black hole,
$r=r_{\pm}=M\pm(M^2-a^2)^{1/2}$. This is the well known {\em
  coordinate} singularity of the black hole metrics.

A coordinate transformation given by
\begin{eqnarray}
d\tilde{\phi} & = & d\phi - \frac{a}{\Delta} \epsilon  dr , 
\label{kerrcoor1} \\
d\tilde{t} &=  & dt - \epsilon
\left[ \frac{1+Y}{1+Y-Z} - \frac{1-Z^k}{1-Z} \right] dr,
\label{kerrcoords}
\end{eqnarray}
where $Y=a^2\sin^{2}\theta/\varrho^2$, $Z=2Mr/\varrho^2$, $k$ a
non-negative integer and $\epsilon=-1(+1)$, regularizes the future
(past) horizon of a rotating black hole (PF98).
With the above ansatz, the metric~(\ref{blform}) becomes, 
in the new coordinates $(\tilde{t},r,\theta,\tilde{\phi})$,
\begin{eqnarray}
ds^2 & = & -(1-Z) d\tilde{t}^2 + \varrho^2 d\theta^2 
+ \sin^2 \theta \varrho^2 (1+ Y (1+Z)) d\tilde{\phi}^2 
+\frac{Z^{2k}-1}{Z-1} dr^2  \nonumber \\ 
& & - 2 a \sin^2 \theta Z d\tilde{t} d\tilde{\phi} 
- 2 \epsilon Z^{k} d\tilde{t} dr 
+ 2 a \epsilon \sin^{2} \theta \frac{Z^{k+1}-1}{Z-1} d\tilde{\phi}dr.
\end{eqnarray}
This form of the metric is now regular at the horizon for any choice
of $k$.

The \{3+1\} decomposition (see, e.g., Misner, Thorne and Wheeler 1973) of this
form of the metric leads to a {\em spatial 3-metric} with non-zero
elements given by
\begin{eqnarray}
\gamma_{rr} & = & \frac{Z^{2k}-1}{Z-1} \\
\gamma_{r\tilde{\phi}} & = & a \epsilon \sin^{2}\theta \frac{Z^{k+1}-1}{Z-1} \\
\gamma_{\theta\theta} & = & \varrho^2 \\
\gamma_{\tilde{\phi}\tilde{\phi}} 
& = & \varrho^2 \sin^2 \theta (1+ Y (1+Z)).
\end{eqnarray}
The components of the {\em shift vector} are given by
\begin{equation}
\beta_{i} = \left( -\epsilon Z^k , 0 , - a \sin^2\theta Z \right),
\end{equation}
and the {\em lapse function} is given by 
\begin{equation}
\alpha^2 = \frac{Z-1}{Z^{2k} - 1 - Y Z^2 (Z^{k-1}-1)/(Z-1) }.
\end{equation}

The form of the \{3+1\} quantities illustrates that for coordinate systems
regular at the horizon, the 3-metric acquires a non-zero off-diagonal
term, whereas the shift vector acquires a radial component. 

The case $k=1$ corresponds to the so called KS form of the Kerr metric
in which the line element reads

\begin{eqnarray}
ds^2 &=& - \left(1-\frac{2Mr}{\varrho^2}\right) d\tilde{t}^2
-\frac{4Mar}{\varrho^2}\sin^2\theta d\tilde{t}d\tilde{\phi} +
\frac{4Mr}{\varrho^2} d\tilde{t} dr +
\left(1+\frac{2Mr}{\varrho^2}\right) dr^2 -
\nonumber \\
& &
2a\left(1+\frac{2Mr}{\varrho^2}\right)\sin^2\theta dr d\tilde{\phi} +
\varrho^2 d\theta^2 +
\sin^2\theta \left[{\varrho^2} + a^2\left(1+\frac{2Mr}{\varrho^2}\right)
\sin^2\theta \right] d\tilde{\phi}^2,
\label{ksform}
\end{eqnarray}
and the corresponding \{3+1\} quantities, considerably simpler than in the
general case, read
\begin{eqnarray}
\gamma_{rr} & = & Z+1  \\
\gamma_{r\tilde{\phi}} & = & a \epsilon \sin^{2}\theta (Z+1) \\
\gamma_{\theta\theta} & = & \varrho^2 \\
\gamma_{\tilde{\phi}\tilde{\phi}} 
& = & \varrho^2 \sin^2 \theta (1+ Y (1+Z)) d\tilde{\phi}^2 \\
\beta_{i} & = & ( -\epsilon Z , 0 , - a \sin^2\theta Z ) \\
\alpha^2 & = & 1/(Z+1).
\end{eqnarray}

It has been argued in PF98 and FIP98 that numerical computations of
matter flows in black hole spacetimes benefit from the use of systems
regular at the horizon. At the same time, the simplicity of the BL
metric element has led to a large body of intuition and the
development of tools based on that system (e.g., to describe the
appearance of accretion disks near the horizon). Hence, it appears useful to
establish the framework for connecting results and simulations in the
two different computational approaches.  This is in general possible
with the appropriate use of the explicitly known coordinate
transformations (Eqs.~(\ref{kerrcoor1}) and ~(\ref{kerrcoords})). A
very important special case occurs for hydrodynamic flows that become,
eventually, stationary. The stationarity permits to {\em map} the
solutions onto the same physical {\em absolute} space. Integrating the
transformation~(\ref{kerrcoor1}) we obtain the angular coordinate of a
given physical point in the two coordinate systems,
\begin{eqnarray}
\phi=\tilde{\phi}-\frac{a}{2\sqrt{M^2-a^2}}
\log\left(\frac{r-M-\sqrt{M^2-a^2}}{r-M+\sqrt{M^2-a^2}}\right).
\end{eqnarray}

In order to compare the velocity components between the two systems
we first transform the components of the fluid 4-velocity, $u^{\mu}$,
according to
\begin{eqnarray}
u^{t} &=& u^{\tilde{t}} - \frac{2Mr}{\Delta} u^{\tilde{r}}
\\
u^{\phi} &=& u^{\tilde{\phi}} - \frac{a}{\Delta} u^{\tilde{r}} \\
u^{r} & = & u^{\tilde{r}} \\
u^{\theta} & = & u^{\tilde{\theta}}. 
\end{eqnarray}
\noindent
Notice that, although the radial and polar components do not change
under the coordinate transformation, we are explicitely using a tilde in $r$
and $\theta$ to indicate KS coordinates. The corresponding transformation
of the Eulerian velocity components is given by
\begin{eqnarray}
v^r &=& \Psi 
\left(v^{\tilde{r}}-\frac{\beta^{\tilde{r}}}{\alpha_{KS}} \right)
\label{transvel1}
\\
v^{\theta} &=& \Psi v^{\tilde{\theta}}
\label{transvel2}
\\
v^{\phi} &=& \Psi
\left(v^{\tilde{\phi}} -\frac{a}{\Delta}v^{\tilde{r}} \right)
- \Psi
\left(\frac{\beta^{\tilde{\phi}}}{\alpha_{KS}} -
\frac{a}{\Delta} \frac{\beta^{\tilde{r}}}{\alpha_{KS}} \right)
+ \frac{\beta^{\phi}}{\alpha_{BL}},
\label{transvel3}
\end{eqnarray}
\noindent
where  $v^{r}$, $v^{\theta}$, and $v^{\phi}$
are related to the proper velocity of the fluid according to
\begin{eqnarray}
v^i = { {u^i} \over {\alpha u^t}} + { {\beta^i} \over {\alpha} },
\end{eqnarray}
\noindent
and $\Psi$ is the ratio between the Lorentz factors, at a given
physical point, in the two coordinate systems, given by
\begin{eqnarray}
\Psi\equiv\frac{W_{BL}}{W_{KS}}=\frac{\alpha_{BL}}{\alpha_{KS}}-
\frac{2Mr}{\Delta}\alpha_{BL}\left(v^{\tilde{r}}-
\frac{\beta^{\tilde{r}}}{\alpha_{KS}} \right).
\label{lorenzratio}
\end{eqnarray}
\noindent
The mapping between the two coordinate systems necessarily breaks down
at the horizon. Hence comparisons will be restricted to some exterior
domain. Similar expressions give the {\em inverse map}.

\subsection{Hydrodynamic equations}

Following the general approach laid out in Banyuls et al. (1997), we
now restrict the domain of integration of the hydrodynamic
equations to the equatorial plane of the Kerr
spacetime, $\theta=\pi/2$. In so doing they adopt the balance law form

\begin{eqnarray}
 {{\partial {\bf U}({\bf w})} \over {\partial t}} +
 {{\partial (\alpha {\bf F}^{r}({\bf w}))} \over {\partial r}} +
 {{\partial (\alpha {\bf F}^{\phi}({\bf w}))} \over {\partial \phi}}
 = {\bf S}({\bf w}),
\label{system}
\end{eqnarray}
\noindent
where $\alpha$ is, again, the lapse function of the spacetime, defined as
$\alpha^2\equiv -1/g^{tt}$.
In Eq.~(\ref{system}) the vector of {\it primitive variables} is
defined as
\begin{eqnarray}
{\bf w} = (\rho, v_{r}, v_{\phi}, \varepsilon)
\end{eqnarray}
\noindent
where $\rho$ and $\varepsilon$ are, respectively, the rest-mass density
(not to be confused with the geometrical factor $\varrho$)
and the specific internal energy, related to the pressure
$p$ via an equation of state which we chose to be that of an ideal gas 
\begin{eqnarray}
p=(\gamma - 1) \rho \varepsilon,
\end{eqnarray}
\noindent
with $\gamma$ being the constant adiabatic index.  

On the other hand,
the vector of unknowns (evolved quantities) in Eq.~(\ref{system}) is
\begin{eqnarray}
{\bf U}({\bf w})  =   (D, S_r, S_{\phi}, \tau).
\end{eqnarray}
\noindent
The explicit relations between the two sets of variables, ${\bf U}$ and
${\bf w}$, are
\begin{eqnarray}
D &=&  \rho W
\nonumber
\\
S_j &=&  \rho h W^2 v_j \,\, (j=r,\phi)
\label{elf}
\\
\tau &=&  \rho h W^2 - p - D,
\nonumber
\end{eqnarray}
\noindent
with $W$ being the Lorentz factor, $W\equiv\alpha u^t =
(1-v^2)^{-1/2}$, with $v^2=g_{ij}v^i v^j$.  The specific form of the
fluxes, ${\bf F}^{i}$, and the source terms, {\bf S}, are given in the
Appendix for the two different representations of the Kerr metric we
use. Further details about the equations can be found in Banyuls et al
(1997).

\subsection{Numerical issues}

We solve system~(\ref{system}) on a discrete numerical grid. To this
end, we take advantage of the explicit hyperbolicity of the system in
order to build up a linearized Riemann solver. Schemes using
approximate Riemann solvers are based on the characteristic
information contained in the system of equations. With this strategy,
physical discontinuities appearing in the solution, e.g., shock waves,
are treated consistently (shock-capturing property).  An explicit
formulation of our numerical algorithm can be found in FI98a. Tests of
the code can be found in Font et al (1994) and Banyuls et al (1997).
We note that in the present work we are using the eigenfields reported
in Font et al (1998), as they extend those presented in Banyuls et al
(1997) to the case of {\it non-diagonal} spatial metrics (such as the
KS line element adopted here).

When using the BL coordinates we choose the inner radius of the
computational domain {\it sufficiently close} to the horizon. In our
computations we have found that the code was very sensitive to this
location, not allowing in some cases proximity to $r_{+}$ without
numerical inaccuracies. The particular values we chose in our
simulations for the inner and outer radial boundaries are summarized
in Table 1.

Resolution requirements near the horizon motivate the use of a
logarithmic radial coordinate for the discretization. This is the well
known {\it tortoise} coordinate, $r_{\star}$, which is defined by
$dr_{\star}=(r^2+a^2) dr/\Delta$, where $r$ is the radial BL
coordinate. This choice permits to use a dense grid of points near the
horizon and has been used extensively in numerical computations (see
e.g., Petrich et al 1989; Abrahams and Shapiro 1990; Krivan et al
1997; FI98a,b). The coordinate singularity of the metric has an
immediate effect on the hydrodynamic quantities. The flow speed
approaches the speed of light and, in consequence, the Lorentz factor
tends to infinity.  As shown by Eq.~(\ref{elf}) this variable couples,
in a non-linear way, the system of equations of relativistic
hydrodynamics. Instabilities caused by extreme behavior of this
quantity lead to the ultimate termination of the code.

We use a typical grid of 200 radial zones and 160 angular zones, in
both coordinate systems. The final accretion pattern is found to be in
the convergence regime already with a coarser angular grid of about 80
zones.  However, we employ a finer angular grid in order to obtain
sharper shock profiles. We only need to impose boundary conditions
along the radial direction. These are the same as those used in
FI98b.  Namely, at the inner boundary we use {\it outflow} conditions,
where all variables are linearly extrapolated to the boundary zones.
At the outer boundary we use the asymptotic initial values of all
variables, for the upstream region, whereas a linear extrapolation is
performed at the downstream region.

\section{Simulations}

\subsection{Initial setup}

As usual in studies of wind accretion the initial models are
characterized by the asymptotic conditions upstream the accretor. We
choose as free parameters the asymptotic velocity $v_{\infty}$, the
sound speed $c_{s_{\infty}}$ and the adiabatic index $\gamma$. The
first two parameters fix the asymptotic Mach number ${\cal
  M}_{\infty}$. Now, we have an additional parameter which is the
specific angular momentum of the black hole, $a$.  The initial models
are listed in Table 1. The first five models in the table describe the
same thermodynamical flow configuration. We use this subset of models
to illustrate the dependence of the flow morphology and accretion
rates on $a$. The black hole spin increases from model 1 (no rotation)
to 5. Note that this last model represents a {\it naked singularity},
as $a>M$. This model has been evolved using only KS coordinates. The
last two models in the table together with model 4 allow us to study
the dependence of our accretion results on the adiabatic index of
the fluid. In these three cases, we consider a rapidly rotating black
hole with $a=0.99M$.  The parameter $a$ is always chosen to be
positive which, in the figures presented below, indicates that the
rotational sense of the black hole is always counter-clockwise.

One of the main aims of the present work is to compare the morphology
of the accretion flow in two different foliations of the black hole spacetime.
In order to properly do this comparison we choose the same
initial (uniform) velocity distribution in both coordinate systems. In 
BL coordinates the velocity field is given by
\begin{eqnarray}
v^{r}(r,\phi) &=& \sqrt{\gamma^{rr}} v_{\infty} \cos{\phi} 
\\
v^{\phi}(r,\phi) &=& -\sqrt{\gamma^{\phi\phi}} v_{\infty} \sin{\phi},
\end{eqnarray}
\noindent
with $\gamma{ij} = g^{ij} + \beta^{i}\beta^{j}/\alpha^2$. Similarly,
in KS coordinates, the initial velocity field, which looks 
algebraically more complicated due to the off-diagonal $g_{r\tilde{\phi}}$
term, reads
\begin{eqnarray}
v^{r}(r,\tilde{\phi}) &=& F_1(r) v_{\infty} \cos{\tilde{\phi}} +
                        F_2(r) v_{\infty} \sin{\tilde{\phi}}
\\
v^{\tilde{\phi}}(r,\tilde{\phi}) &=& -F_3(r) v_{\infty} \sin{\tilde{\phi}} +
                        F_4(r) v_{\infty} \cos{\tilde{\phi}},
\end{eqnarray}
\noindent
with 
\begin{eqnarray}
F_1(r) &=& \frac{1}{\sqrt{g_{rr}}}
\\
F_4(r) &=& -\frac{2g_{r\tilde{\phi}}}{\sqrt{g_{rr}}
                  g_{\tilde{\phi}\tilde{\phi}}}
\\
F_3(r) &=& \frac{F_1 g_{rr} + F_4 g_{r\tilde{\phi}}}
  {\sqrt{ (g_{rr}g_{\tilde{\phi}\tilde{\phi}} - g_{r\tilde{\phi}}^2)
  (F_1^2g_{rr}+F_4^2g_{\tilde{\phi}\tilde{\phi}} +
  2F_1F_4g_{r\tilde{\phi}}) }}
\\
F_2(r) &=& \frac{ F_3F_4g_{\tilde{\phi}\tilde{\phi}} +
                  F_1F_3g_{r\tilde{\phi}} }
                { F_1g_{rr} + F_4g_{r\tilde{\phi}} }.
\end{eqnarray}
\noindent
Both velocity fields guarantee initially $v^2=v_{\infty}^2$.

\subsection{Flow morphology}

In Fig.~1 we display the flow pattern for the first four models of
Table 1, which corresponds to a final time of $500M$. This is a
Cartesian plot where the $x$ and $y$ axis are, respectively,
$r\cos{\phi}$ and $r\sin{\phi}$.  We plot isocontours of the logarithm
of the rest-mass density, properly scaled by its asymptotic value.
These results are obtained employing BL coordinates.  The outer domain
of this figure corresponds to $20M$.  Similarly, in Fig.~2 we plot a
close-up view of that domain, up to a distance of $4M$.  The dotted
innermost line, in both figures, represents the location of the
horizon $r_+$.

All models are characterized by the presence of a well-defined tail
shock. As already shown in FI98b, in non-axisymmetric relativistic
Bondi-Hoyle accretion simulations, this shock appears stable to
tangential oscillations, in contrast to Newtonian simulations with
tiny accretors (see, e.g., Benensohn et al 1997 and references there
in). By direct inspection of these figures, the effect of the rotation
of the black hole on the flow morphology becomes clear. The shock
becomes wrapped around the central accretor as the black hole angular
momentum $a$ increases. The effect is, though, localized to the
central regions. In the $a=0.5$ plot we see that the morphology of the
shock deviates from the pattern of the non-rotating case inside
approximately $r=4M$. For $a=0.9$, the region of influence extends
slightly further out, to about $r=5M$.  At the outer regions, as
expected, the overall morphology is remarkably similar for all values
of $a$.  This is not surprising, since the Kerr metric rapidly
approaches the Schwarzschild form for radii large compared to
$r_{+}$. This fact, in turn, shows that possible telltale signals of a
rotating black hole,in an astrophysical context, demand an accurate
description of the innermost regions.

In Figs.~3 and 4 we show the final configuration for the density for
the same initial setups as in Figs.~1 and 2, but now for simulations
employing the KS coordinate system.  We construct the plots using the
standard transformation between KS and Cartesian coordinates $x + i y
= (r + i a) \sin\theta e^{i \tilde{\phi}}$ (Hawking and Ellis 1972)
with $\theta=\pi/2$ .  Again, the dotted line in these plots indicates
the position of $r_{+}$.  Notice that in KS coordinates the
computation extends inside the horizon (this is more clearly seen in
Fig.~4). The flow morphology shows smooth behaviour when crossing the
horizon, all matter fields being regular there.  We observe here the
same confinement of rotational effects to the innermost regions around
the black hole. Notably though, the shock structure appears less
deformed, especially the lower component.  The reason is that in the
BL description of the black hole geometry, the dominant effects near
the horizon are purely kinematic (associated with coordinate system
pathologies) and dissapear with the adoption of a regular
system. Clearly, the accurate description of near horizon effects can
be achieved easier when using horizon adapted coordinate systems.

As in the non-rotating studies (FI98a,b), we also notice that the most
efficient region of the accretion process is the rear part of the
black hole. The material flowing inside some radius smaller than the
characteristic impact parameter of the problem (say, the accretion
radius, see below) after changing its sense of motion due to the
strong gravitational field is acummulated behind the black hole and is
gradually accreted. We find that this maximum density always increases
with $a$ and its functional dependence is clearly greater than linear.
The specific values we obtain can be found in the corresponding
captions of Figs.~1 to 4. Typical density enhancements in the
post-shock region (BL coordinates) with respect to the asymptotic
density range in between $1.65$ ($a=0$) and $1.77$ ($a=0.99$) (in
logarithmic scale).

Now we turn to the description of the flow morphology for different
values of $\gamma$, the fluid adiabatic exponent. The accretion
patterns for models 6 ($\gamma=4/3$), 4 ($\gamma=5/3$) and 7
($\gamma=2$) are depicted in Fig.~5. Once more, the variable we show
in this figure is the logarithm of the scaled rest-mass density.
Clearly visible in this plot are the larger shock opening angles for
the larger values of $\gamma$. This is explained by the enhanced
values of the pressure inside the shock ``cone" as $\gamma$ increases.
We already noticed this behaviour in the non-rotating simulations
performed in FI98a,b. Now, the larger values of $\gamma$, combined
with the rapid rotation of the black hole ($a=0.99$), wrap the upper 
shock wave around the accretor. This effect is more pronounced for the 
larger $\gamma$ values. We also note that the lower shock wave is less
affected by the increase in $\gamma$. While it still opens to larger
angles, the existing rotational flow counteracts the effects of the
pressure force keeping its position almost unchanged.  

The enhancement of the pressure in the post-shock zone is responsible
for the so-called ``drag'' force experienced by the accretor.  We
notice here that the rotating black hole is redistributing the high
pressure area, with non-trivial effects on the nature of the drag
force. Whereas in the Schwarzschild case the drag force is alligned
with the flow lines, pointing in the upstream direction, in the Kerr
case we notice a distinct asymmetry between the co-rotating and
counter-rotating side of the flow. The pressure enhancement is
predominantly on the counter-rotating side. In Fig.~6 this
observation is made more precise with the examination of the pressure
profile, at the innermost radius, for the $a=0.99$ case.  Three
different $\gamma$ values are illustrated, showing the strong
dependence of the pressure asymmetry on the adiabatic index.  This is
particularly clear in the limiting case $\gamma=2$ (dashed line). We
observe a pressure difference of almost two orders of magnitude, along
the axis normal to the asymptotic flow direction.  The implication of
this asymmetry is that a rotating hole moving accross the interstellar
medium (or accreting from a wind), will experience, on top of the drag
force, a ``lift'' force, normal to its direction of motion (to the
wind direction). 

It is interesting to note that this effect bears a strong superficial
resemblance to the so-called ``Magnus'' effect, i.e., the experience
of lift forces by rotating bodies immersed in a stream flow. There,
the lift force is due to the increased speed of the flow on the
co-rotating side (due to friction with the object), and the increase
of pressure on the counter-rotating side (which follows immediately
from the Bernoulli equation). We note that the direction of the lift,
in relation to the sense of rotation, agrees in both contexts. We
caution though that the underlying causes may be very different. In
the black hole case the flow is supersonic and there is no boundary
layer.

Completing the study of the broad morphology of the flow and its
dependence on the black hole spin, we extend the value of $a$ above
$M$, choosing, in particular, $a=1.1M$ (model 5). This case
corresponds to accretion onto a {\it naked singularity}. Although from
the theoretical point of view such objects are believed not to exist
in Nature (according to the {\it Cosmic Censorship hypothesis}, all
{\it physical} singularities formed by the gravitational collapse of
nonsingular, asymptotically flat initial data, must be hidden from the
exterior world inside an event horizon) we nonetheless decided to
perform such a computation, in order to assess the behaviour of the
code in this regime, and to explore the extrapolation of previous
simulations.  The resulting morphology for this simulation, using KS
coordinates, is plotted in Fig.~7. We are showing isocontours of the
logarithm of the rest-mass density in a region extending $4M$ in the
$x$ and $y$ directions from the singularity. In this situation, there
is an ambiguity as to where to place the inner boundary of the domain.
The closer one gets to $r=0$ the stronger the gravity becomes (with
infinite tidal forces at the singularity). This introduces important
resolution requirements on the numerical code. For this reason we
chose $r_{min}=M$, in accordance with the location of the inner
boundary in the maximal case $a=M$. As can be seen from Fig.~7 the
flow morphology for this model follows the previous trend found for
lower values of $a$ (models 1 to 4): the shock appears slightly more
wrapped (around the $r=M$ circle) and the maximum rest-mass density in
the rear part of the accretor increases.

\subsection{Accretion rates}

We compute the accretion rates of mass, radial momentum and angular
momentum. The procedure of computation can be found in FI98a and
FI98b. The rates are typically evaluated at the accretion radius,
$r_a$, defined as
\begin{eqnarray}
r_a=\frac{M}{v_{\infty}^2 + c_{s_{\infty}}^2}.  
\end{eqnarray}
For our models $r_a=3.85M$. The results are plotted in Figs.~8-10 for
BL coordinates and in Figs.~11-13 for the KS system.  In these figures
we show the time evolution of those rates for the whole simulation.
For comparison purposes, the radial and angular momentum accretion
rates have been scaled to one (scaling by a factor of 1250 and 80,
respectively in the BL coordinate runs; correspondingly, with KS
coordinates these factors are 300 and 400). We have checked that, as
expected, the qualitative behavior of the accretion rates is
independent of the radius at which they are computed, as long as a
stationary solution is found.

All rates show a clear transition to a final stationary state, around
the time interval $100M-200M$, regardless of the coordinate system
used. The non-rotating case ($a=0$) shows no signs of oscillations,
leading to remarkably constant values for all rates (again independent
of the coordinates). Non zero $a$ values are seen to show considerable
more oscillatory behavior around some average value. The reason behind
this effect is purely numerical. The black hole geometry leads, in the
rotating case, to a significantly larger number of non-zero
Christoffel symbols, which appear explicitly in the source terms of
the hydrodynamic equations (see the Appendix). Increasing the
resolution (in particular the angular one), considerably reduces the
amplitude of the oscillations, as can be read off Fig.~14.

The mass accretion rates have been scaled to the canonical value
proposed by Petrich et al (1989). In KS coordinates (Fig.~11) this
value is larger than in BL coordinates (Fig.~8). However, this is just
a coordinate effect as we show in the next section. In addition, in
Fig.~8 there appears to be a certain trend on the variation of the
mass rate with $a$, being (slightly) larger for larger values of $a$.
This is not the case for KS coordinates, where the normalized mass
accretion rate is around $156$, regardless of the value of $a$.  We
will come back to this issue later in this section.

In Fig.~9 we plot the (scaled) radial momentum accretion rate for BL
coordinates. The maximum drag rate is obtained for non-rotating holes.
As for the mass accretion rate, the radial momentum rate also shows a
clear dependence on $a$, especially when using BL coordinates. This 
dependence is not so clear for KS coordinates (see Fig.~12).
This discrepancy will be explained below. Note also that
due to the different scale factors used, the radial momentum
rates in Fig.~9 are a factor 4 larger than those of Fig.~12. 

Finally, the angular momentum accretion rates (Figs.~10 and 13) are
clearly non-zero for rotating black holes in contrast to the $a=0$
case. For both coordinate systems it exhibits a clear trend: the
larger the value of $a$ the larger the angular momentum rate. Due to
the different scales used in the different coordinates, the angular
momentum rates in Fig.~10 are a factor 5 smaller than those in Fig.~13.

To clarify the discrepancies found in the computation of the accretion
rates in the different coordinate systems we plot in Fig.~15 the
dependence of those rates with the specific angular momentum of the
black hole. In order to do so, we average the rates over the final
$200M$ of the evolution, once the steady state is well established.
Results for the BL system are depicted with a filled triangle whereas
results for the KS system are represented by filled circles. We are
only considering the subset of models 1-5 in Table 1. As in the
previous figures, the momentum accretion rates have been properly
scaled with adequate factors. The radial momentum rates presented in
this plot are computed both at $r_a$ (solid lines) and at $r_{max}$
(dashed lines). From this figure the non-dependence of the mass
accretion rate on $a$ becomes more clear.  This is the expected
result, as the mass accretion rate is an integral invariant, and hence
does not change by deformations of the surface over which it is
computed. Deforming this surface outwards, by increasing the radius at
which the mass accretion rate is measured, must lead to the same
result, so long as the solution is in a steady state. At sufficiently
large radii, the rotating aspect of the metric is essentially
``hidden" and the accretion rate should depend only on the total mass
of the central potential. We conclude that the larger discrepancies
found when computing the mass accretion rate in the BL coordinates
(see Fig.~7 for models 1 to 4) are purely due to numerical reasons,
induced for instance, by the more extreme dynamical behaviour of some
fields in the near zone of the black hole potential.  However, the
deviations never exceed a few percent, more specifically, $1.5\%$ for
the BL system simulation and $1.2\%$ using KS coordinates.

Coming back to the radial momentum accretion rate computation, we have
verified that its dependence on the black hole angular momentum is
strongly related to the radius at which it is computed.  We find that
at large radii its value remains remarkably constant for all values of
$a$. In Figs.~9 and 12 we plot the drag rate computed at the accretion
radius. In BL coordinates (Fig.~9) we get a $15\%$ difference between
the $a=0$ and $a=0.99$ cases. For KS coordinates (Fig.~12) this
difference is reduced to $5\%$. However, if the radial momentum
accretion rate is computed at $r_{max}$ (as in Fig.~15, dashed lines)
we obtain significantly smaller differences: $7\%$ for BL coordinates
and only $0.4\%$ for KS coordinates.

We quantify next how the mass accretion rate depends on the arbitrary
location (but necessarily larger than $r_+$) of the inner boundary in
BL coordinates. Placing this inner boundary at $r=2.1M$ for model 3 we
find the surprising result that, although the broad flow morphology
looks qualitatively similar in both cases, the accretion rates {\it
strongly} depend on the value of the innermost radius. We note that a
small change in location (such as from $1.89M$ to $2.1M$) introduces,
roughly, a $10\%$ difference on the computed mass accretion rate.

Fig.~15 also illustrates the dependence of the angular momentum
accretion rate with the spin of the black hole. Clearly, this quantity
increases with $a$ in a non-linear way.

We complete our study of the different accretion rates presenting
their dependence on the fluid adiabatic index. To do this, we choose the
rapidly rotating hole ($a=0.99M$) of models 4, 6 and 7. Table 2 shows
the results obtained using the KS coordinate system. Again, we are
averaging the accretion rates on the final $200M$ of evolution.  We
note that we are now choosing a different normalization. The reason is
that the canonical mass accretion rate we are using to scale all rates
(see Petrich et al 1989; see also Shapiro and Teukolsky 1983) is ill
defined for $\gamma >5/3$. As we are mostly interested in their
qualitative behaviour and not in the particular numbers,
re-normalizing all rates to unity suffices for our purposes. Our
numerical study shows that all accretion rates decrease (in absolute
value for the momentum rates) as $\gamma$ increases from $4/3$ to $2$.
This is in contrast with the results presented in FI98a for the
non-rotating case. However, this is just an apparent disagreement
related to the different normalization procedures employed in the two
surveys.

\subsection{Coordinate system comparison}

We focus now on a direct comparison between the accretion patterns
obtained with the different coordinate systems. In order to do so we
take advantage of the stationarity of the solution at late times.  In
Fig.~\ref{mapping} we show the setup that allows for a simple
comparison between the BL and KS coordinate systems in the case of
stationary flows. For such flows, a one-to-one correspondence between
physical points can be established using the appropriate coordinate
transformations presented before.

The procedure of comparison involves a transformation from
$(r,\tilde{\phi})$ coordinates to $(r,\phi)$ and, finally, to
$(x,y)\equiv(r\cos\phi,r\sin\phi)$. For the case of scalar quantities,
such as the density, the comparison procedure ends here.  For vector
fields, such as the velocity, we employ in addition the linear
transformations given by Eqs.~(\ref{transvel1})-(\ref{transvel3}).

We focus on model 4 of Table 1 ($a=0.99M$). Fig.~17 shows the
isodensity contours at the final time ($t=500M$) in BL coordinates.
The left panel corresponds to an actual BL evolution, while the right
one shows how results from a simulation originally performed in KS
coordinates look like when transformed to BL coordinates. The
innermost dotted circle marks the location of the horizon. The dashed
line marks the position of $r_{min}$ in BL coordinates (we
intentionally cut out the interior region in the right panel). 
The qualitative agreement of the two plots is remarkable.

We turn now to the comparison of the radial and azimuthal components
of the velocity. This is depicted in Fig.~18. The convention for the
left and right panels follows the same criteria as in Fig.~17. Again,
the agreement is excellent, even though comparing these two quantities
involves, besides the coordinate shift (from $\tilde{\phi}$ to
$\phi$), also a non-trivial linear transformation.

From these figures one can notice that the smooth features of the
solution (i.e., the pre-shocked part of the flow) are well captured in
both systems, whereas the shock and the innermost regions present
slightly more numerical noise in BL coordinates. Notice also that,
although in BL coordinates we are forced to cut out an important
domain of integration (due to numerical instabilities associated to
this pathological system), it is also true that the solution does not
seem to be qualititavely affected, at least for the flows considered
here. This may be explained by the fact that the flow is highly
supersonic at the innermost zone. This simplifies the task of imposing
appropriate boundary conditions there, e.g., a simple linear
extrapolation will always work.

It is instructive to see how the solution would look like in BL
coordinates if the location of $r_{min}$ were moved inwards.  This is
plotted in Fig.~19. Here we show isodensity contours starting in a
region much closer to the horizon than in Fig.~17 (note that at the
horizon the coordinate transformation is singular). We can now follow
the shock location all the way down to those inner regions, showing
the singular spiralling around the central hole.

As our final comparison between quantities computed in the two
coordinate systems we study the correlation of the mass accretion
rates.  In order to do so, we must compute the mass rate with respect
to {\it proper} time, through a given physical surface.  The relation
between the proper, $\tau$, and coordinate, $t$, times (which is the
one employed in the plots of the accretion rates) involves the lapse
function, $d\tau=\alpha dt$. Making the comparison, hence, requires
including the ratio of the lapse in the two coordinate systems at that
physical location. The final relation is given by the surface integral
\begin{eqnarray}
{\dot m}_{BL} = -\int
\frac{1}{\Psi} D \alpha \sqrt{\gamma}
\left(v^{\tilde r}-\frac{\beta^{\tilde{r}}}{\alpha}\right) d\Sigma
\end{eqnarray}
\noindent
where ${\dot m}$ denotes the rate of change in (coordinate) time of
the mass accreted and all quantities (except $\Psi$, see
Eq.~(\ref{lorenzratio})) in the integrand are computed in the KS
system. The result of the comparison, for model 4, is plotted in Fig.~20. 
We show the mass accretion rate sampled at discrete (coordinate) times (every
$50M$). The circles indicate the mass accretion estimate in the {\it
  original} BL simulation. The plus signs show this rate as a result
of the transformation from the KS simulation.  We can again conclude
that the qualitative agreement is good, especially, and as expected
by the comparison procedure we use, once the steady-state is reached.

\section{Discussion}

In this paper we have presented detailed numerical computations of
non-axisymmetric relativistic Bondi-Hoyle accretion onto a rotating
(Kerr) black hole. The integrations have been performed with an
advanced {\it high-resolution shock-capturing} scheme based on an
approximate Riemann solver. In particular, we have studied accretion
flows onto rapidly-rotating black holes.

We have demonstrated that even in the presence of rotation, the
relativistic accretion patterns always proceed in a stationary
way. This seems to be a common feature of relativistic flows as
opposed to non-axisymmetric Newtonian computations. Previous
relativistic simulations for Schwarzschild (non-rotating) black holes
already pointed out this stability (FI98b).  The {\it physical}
minimum size of the accretor considered now is $r_+=M$, the
(outermost) event horizon of a Kerr black hole. For our initial data,
the minimum value corresponds to $0.29 r_a$, with $r_a$ being the
accretion radius. This parameter controls the appearance of the
tangential instability in Newtonian flows.  The instability was found
there to appear only for {\it tiny} accretors.  Although we have now
smaller accretors than in the Schwarzschild case the flow patterns
still relax to a final steady-state.

The effects of the black hole rotation on the flow morphology were
seen to be confined to the inner regions of the black hole
potential. Within this region, the black hole angular momentum drags
the flow, wrapping the shock structure around, and generating an
overpressure on the counter-rotating side. This is reminiscent of
similar behavior in classical fluid mechanics (i.e., the Magnus
effect), although there does not appear to be a deeper physical
similarity between the two contexts.

As a hypothetical scenario, we have also considered accretion flows
onto a naked singularity ($a>M$). Our preliminary observation is that
the morphology of the flow in this case is a smooth continuation of
the black hole simulations.

The validity of the results has been double-checked by performing the
simulations in two different coordinate systems.  A gratifying result
of our study, confirming the accuracy of the computations, is the
overall agreement obtained in this comparison (performed for
a variety of matter fields), taking advantage of the stationarity of
the final accretion flow.  Although the transformations of scalars and
vectors from one coordinate system to the other are far from trivial,
we have found good overall agreement in our results.

The stationarity of the solution has been also demonstrated computing
the accretion rates of mass and momentum. Those rates were found to
depend on the coordinate system used, but they roughly agree when
transformed to the same frame.  The mass and radial momentum rates
show (slight) dependence on the spin of the black hole when using BL
coordinates, but not so for the KS system. As those quantities should
be independent of the black hole angular momentum, the computations
using KS coordinates were, numerically, more accurate.  On the other
hand, as one would expect, the angular momentum accretion rates vanish
only for Schwarzschild holes and substantially increase as $a$
increases.  We have also presented the dependence of the different
accretion rates on the adiabatic index of the (perfect fluid) equation
of state. The results found for a rotating black hole with $a=0.99M$
show smaller values for all rates as $\gamma$ increases.

We have shown that the choice of the Kerr-Schild form of the Kerr
metric, regular at the horizon, allows for more accurate integrations
of the general relativistic hydrodynamic equations than the {\it
standard} singular choice (i.e., Boyer-Lindquist coordinates).  Such
horizon adapted (regular) systems eliminate numerical inconsistencies
by placing the inner boundary of the domain inside the black hole
horizon, hence, causally disconnecting unwanted boundary influences.

\section*{Acknowledgments}

This work has been partially supported by the Spanish DGICYT (grant
PB97-1432) and the Max-Planck-Gesellschaft.  
J.A.F. has benefited from a TMR European contract
(nr. ERBFMBICT971902). P.P. would like to thank SISSA for warm
hospitality while part of this work was completed. J.M.I. thanks
S. Bonazzola and B. Carter for encouraging comments and
interesting suggestions.  We also want to thank M. Miller for notifying us
an erratum found in the eigenvectors reported in Banyuls et al (1997)
which are only strictly valid for {\it diagonal} spatial metrics.
All computations have been performed at the Albert Einstein Institute in 
Potsdam.

\appendix

\section*{Appendix}
\label{appendix}

In this appendix we write down, explicitly, the fluxes and the source
terms of the general relativistic hydrodynamic equations for the Kerr
line element.  In addition, we also write the non-vanishing
Christoffel symbols needed for the computation of the sources. All the
expressions are written for the two different system of coordinates
used in the computations and are specialized for the equatorial plane
$(\theta=\pi/2)$.

\vspace{0.2cm}
\noindent
\begin{center}
{\bf Fluxes}
\end{center}

\vspace{0.2cm}
\noindent
{\it Boyer-Lindquist}:

\begin{eqnarray}
{\bf F}^{r}({\bf w})  =   \left(D v^{r},
 S_r v^{r} + p, S_{\phi} v^{r},
(\tau + p) v^{r} \right)
\end{eqnarray}
\noindent
\begin{eqnarray}
{\bf F}^{\phi}({\bf w})  =   \left(D
   \left(v^{\phi} - { {\beta^{\phi}} \over {\alpha} }\right),
S_{r} \left(v^{\phi} - { {\beta^{\phi}} \over {\alpha} }\right), 
S_{\phi} \left(v^{\phi} - { {\beta^{\phi}} \over {\alpha} }\right) + p,
\tau \left(v^{\phi} - { {\beta^{\phi}} \over {\alpha} }\right) + 
p v^{\phi} \right).
\end{eqnarray}
\noindent

\vspace{0.2cm}
\noindent
{\it Kerr-Schild}:

\begin{eqnarray}
{\bf F}^{r}({\bf w})  =   \left(D \left(v^{r}-\frac{\beta^r}{\alpha} \right),
S_r \left(v^{r}-\frac{\beta^r}{\alpha} \right) + p, 
S_{\tilde{\phi}} \left(v^{r}-\frac{\beta^r}{\alpha} \right),
\tau \left(v^{r}-\frac{\beta^r}{\alpha} \right) + p v^{r} \right)
\end{eqnarray}
\noindent
\begin{eqnarray}
{\bf F}^{\tilde{\phi}}({\bf w})  =   \left(D v^{\tilde{\phi}},
 S_r v^{\tilde{\phi}}, S_{\tilde{\phi}} v^{\tilde{\phi}} + p,
(\tau + p) v^{\tilde{\phi}} \right)
\end{eqnarray}

\vspace{0.2cm}
\noindent
\begin{center}
{\bf Sources}
\end{center}

\begin{eqnarray}
{\bf S}({\bf w}) = (S_1,S_2,S_3,S_4)
\end{eqnarray}

\vspace{0.2cm}
\noindent
{\it Boyer-Lindquist}:

\begin{eqnarray}
S_1 & = & -\alpha D v^r \Omega 
\\
S_2 & = & -\alpha (S_r v^r + p) \Omega +
 \alpha T^{rr} g_{rr,r} - \alpha A_1 g_{rr} 
\\
S_3 & = & -\alpha S_{\phi} v^r \Omega +
 \alpha T^{rt} g_{t\phi,r} + \alpha T^{r\phi} g_{\phi\phi,r} -
2 \alpha A_2 g_{t\phi} -
2\alpha A_3 g_{\phi\phi} 
\\
S_4 & = & -\alpha (\tau + p) v^r \Omega + \alpha T^{tr} \alpha_{,r} -
2\alpha^2 A_2
\end{eqnarray}
\noindent
with the definitions
\begin{eqnarray}
\Omega &=& { {1} \over {r} } +
         { {\Sigma_{,r}} \over {2\Sigma} } -
         { {\Delta_{,r}} \over {2\Delta} }
\\
A_1 &=& T^{tt} \Gamma^{r}_{tt} +
 T^{rr} \Gamma^{r}_{rr} + T^{\theta\theta} \Gamma^{r}_{\theta\theta} +
 T^{\phi\phi} \Gamma^{r}_{\phi\phi} + 2 T^{t\phi} \Gamma^{r}_{t\phi}
\\
A_2 &=& T^{r\phi} \Gamma^{t}_{r\phi} + T^{tr} \Gamma^{t}_{tr}
\\
A_3 &=& T^{r\phi} \Gamma^{\phi}_{r\phi} + T^{tr} \Gamma^{\phi}_{tr}
\end{eqnarray}
\noindent
with $T^{\mu\nu}$ being the perfect fluid stress-energy tensor
\begin{eqnarray}
T^{\mu\nu} &=& \rho h u^{\mu} u^{\nu} + p g^{\mu\nu}.
\end{eqnarray}
\noindent
The `,' denotes partial differentiation and the $\Gamma^{\delta}_{\mu\nu}$
stand for the Christoffel symbols. Quantity $h$ appearing in the stress-energy
tensor is the specific enthalpy, $h=1+\varepsilon+p/\rho$.

\vspace{0.2cm}
\noindent
{\it Kerr-Schild}:

\begin{eqnarray}
S_1 & = & -\alpha D \left(v^r -\frac{\beta^r}{\alpha}\right) \Theta
\\
S_2 & = & -\alpha \left(S_r \left(v^r-\frac{\beta^r}{\alpha}\right) + 
p\right) \Theta +
\nonumber
\\
& &
\alpha \left[ 
T^{{\tilde t}r} g_{{\tilde t}r,r} + 
T^{rr} g_{rr,r} + 
T^{r{\tilde\phi}} g_{r{\tilde\phi},r} -
\right.
\left.
\beta_r B_1 -
g_{rr} B_2 -
g_{r{\tilde\phi}} B_3 
\right]
\\
S_3 &=& -\alpha \left(S_{{\tilde\phi}} \left(v^r-\frac{\beta^r}{\alpha}\right)
\right) \Theta +
\nonumber
\\
& &
\alpha \left[
T^{{\tilde t}r} g_{{\tilde t}{\tilde\phi},r} +
T^{rr} g_{r{\tilde\phi},r} +
T^{r{\tilde\phi}} g_{{\tilde\phi}{\tilde\phi},r} -
\right.
\left.
\beta_{\tilde\phi} B_1 -
g_{r{\tilde\phi}} B_2 -
g_{{\tilde\phi}{\tilde\phi}} B_3
\right]
\\
S_4 &=& -\alpha \left(\tau\left(v^r-\frac{\beta^r}{\alpha}\right)+
p v^r\right) \Theta +
\alpha T^{{\tilde t}r} \alpha_{,r} -
\alpha^2 B_1
\end{eqnarray}
\noindent
with
\begin{eqnarray}
\Theta &=& \frac{2r-M\alpha^2}{r^2}
\\
B_1 &=& 
T^{{\tilde t}{\tilde t}} \Gamma^{{\tilde t}}_{{\tilde t}{\tilde t}} +
T^{rr} \Gamma^{{\tilde t}}_{rr} +
T^{\theta\theta} \Gamma^{{\tilde t}}_{\theta\theta} +
T^{{\tilde\phi}{\tilde\phi}} \Gamma^{{\tilde t}}_{{\tilde\phi}{\tilde\phi}} +
2(
T^{{\tilde t}r} \Gamma^{{\tilde t}}_{{\tilde t}r} +
T^{{\tilde t}{\tilde\phi}} \Gamma^{{\tilde t}}_{{\tilde t}{\tilde\phi}} +
T^{r{\tilde\phi}} \Gamma^{{\tilde t}}_{r{\tilde\phi}})
\\
B_2 &=&
T^{{\tilde t}{\tilde t}} \Gamma^{r}_{{\tilde t}{\tilde t}} +
T^{rr} \Gamma^{r}_{rr} +
T^{\theta\theta} \Gamma^{r}_{\theta\theta} +
T^{{\tilde\phi}{\tilde\phi}} \Gamma^{r}_{{\tilde\phi}{\tilde\phi}} +
2(
T^{{\tilde t}r} \Gamma^{r}_{{\tilde t}r} +
T^{{\tilde t}{\tilde\phi}} \Gamma^{r}_{{\tilde t}{\tilde\phi}} +
T^{r{\tilde\phi}} \Gamma^{r}_{r{\tilde\phi}})
\\
B_3 &=&
T^{{\tilde t}{\tilde t}} \Gamma^{{\tilde\phi}}_{{\tilde t}{\tilde t}} +
T^{rr} \Gamma^{{\tilde\phi}}_{rr} +
T^{\theta\theta} \Gamma^{{\tilde\phi}}_{\theta\theta} +
T^{{\tilde\phi}{\tilde\phi}} \Gamma^{{\tilde\phi}}_{{\tilde\phi}{\tilde\phi}} +
2(
T^{{\tilde t}r} \Gamma^{{\tilde\phi}}_{{\tilde t}r} +
T^{{\tilde t}{\tilde\phi}} \Gamma^{{\tilde\phi}}_{{\tilde t}{\tilde\phi}} +
T^{r{\tilde\phi}} \Gamma^{{\tilde\phi}}_{r{\tilde\phi}})
\end{eqnarray}

\vspace{0.2cm}
\noindent
\begin{center}
{\bf Non-vanishing Christoffel symbols}
\end{center}

\vspace{0.2cm}
\noindent
{\it Boyer-Lindquist}:

\begin{eqnarray}
\Gamma^{t}_{tr} &=&  { {M} \over {\Delta r^2} } (a^2+r^2)
\\
\Gamma^{t}_{r\phi} &=& - { {aM} \over {\Delta r^2} } (a^2 + 3r^2)
\\
\Gamma^{r}_{tt} &=& - { {\Delta M} \over {r^4} }
\\
\Gamma^{r}_{rr} &=& { {1} \over {r\Delta} } (a^2 - Mr)
\\
\Gamma^{r}_{\theta\theta} &=& - { {\Delta} \over {r}}
\\
\Gamma^{r}_{\phi\phi} &=& { {\Delta} \over {r^4}}
           (Ma^2 - r^3)
\\
\Gamma^{r}_{t\phi} &=& - { {Ma\Delta} \over {r^4} }
\\
\Gamma^{\phi}_{r\phi} &=& - { {1} \over {\Delta r^2} } (Ma^2 + r^2(2M-r))
\\
\Gamma^{\phi}_{tr} &=& { {Ma} \over {\Delta r^2} }
\end{eqnarray}

\vspace{0.2cm}
\noindent
{\it Kerr-Schild}:

\begin{eqnarray}
\Gamma^{\tilde{t}}_{\tilde{t}\tilde{t}} &=& \frac{2M^2}{r^3}
\\
\Gamma^{\tilde{t}}_{\tilde{t}r} &=& \frac{M(r+2M)}{r^3}
\\
\Gamma^{\tilde{t}}_{\tilde{t}\tilde{\phi}} &=&
-\frac{2M^2a}{r^3}
\\
\Gamma^{\tilde{t}}_{rr} &=& \frac{2M(M+r)}{r^3}
\\
\Gamma^{\tilde{t}}_{r\tilde{\phi}} &=&
-\frac{Ma(r+2M)}{r^3}
\\
\Gamma^{\tilde{t}}_{\theta\theta} &=& -2M
\\
\Gamma^{\tilde{t}}_{\tilde{\phi}\tilde{\phi}} &=&
\frac{2M(Ma^2-r^3)}{r^3}
\\
\Gamma^{r}_{\tilde{t}\tilde{t}} &=& \frac{M\Delta}{r^4}
\\
\Gamma^{r}_{\tilde{t}r} &=& 
-\frac{M(2Mr-a^2)}{r^4}
\\
\Gamma^{r}_{\tilde{t}\tilde{\phi}} &=& -\frac{Ma\Delta}{r^4}
\\
\Gamma^{r}_{rr} &=& -\frac{M(2Mr+r^2-a^2)}{r^4}
\\
\Gamma^{r}_{r\tilde{\phi}} &=& \frac{a(2M^2r + r^3 -Ma^2)}{r^4}
\\
\Gamma^{r}_{\theta\theta} &=& -\frac{\Delta}{r}
\\
\Gamma^{r}_{\tilde{\phi}\tilde{\phi}} &=&
-\frac{r^5-2Mr^4+a^2r^3-Ma^2r^2+2M^2a^2r-Ma^4}{r^4}
\\
\Gamma^{\theta}_{r\theta} &=& \frac{1}{r}
\\
\Gamma^{\tilde{\phi}}_{\tilde{t}\tilde{t}} &=& \frac{Ma}{r^4}
\\
\Gamma^{\tilde{\phi}}_{\tilde{t}r} &=& \frac{Ma}{r^4}
\\
\Gamma^{\tilde{\phi}}_{\tilde{t}\tilde{\phi}} &=& -\frac{Ma}{r^4}
\\
\Gamma^{\tilde{\phi}}_{rr} &=& \frac{Ma}{r^4}
\\
\Gamma^{\tilde{\phi}}_{r\tilde{\phi}} &=& -\frac{Ma^2-r^3}{r^4}
\\
\Gamma^{\tilde{\phi}}_{\theta\theta} &=& -\frac{a}{r}
\\
\Gamma^{\tilde{\phi}}_{\tilde{\phi}\tilde{\phi}} &=&
\frac{a(Ma^2-r^3)}{r^4}
\end{eqnarray}

\newpage

%
% TABLES
%

\begin{center}
    \begin{tabular}{|cccccccccc|}
    \multicolumn{10}{c}{\bf Table 1 \rm} \\
    \multicolumn{10}{c}{\bf Initial models\rm} \\
    \multicolumn{10}{c}{}\\
    \hline
      MODEL & 
      $\gamma$ &
      ${\cal M}_{\infty}$ & 
      $v_{\infty}$ &
      $a$ & 
      $r_{+}$ &
      $r_{min}$ (BL) & 
      $r_{max}$ (BL) &
      $r_{min}$ (KS) &
      $r_{max}$ (KS)     \cr
\hline
\hline
      1     & $5/3$  & $5.0$  & $0.5$ & $0$   & $2$
            & $2.2$  & $38.5$ & $1.8$ & $50.9$          \cr
      2     & $5/3$  & $5.0$  & $0.5$ & $0.5$ & $1.87$
            & $2.08$ & $34.7$ & $1.8$ & $50.9$          \cr
      3     & $5/3$  & $5.0$  & $0.5$ & $0.9$ & $1.44$
            & $1.89$ & $34.9$ & $1.4$ & $50.9$          \cr
      4     & $5/3$  & $5.0$  & $0.5$ & $0.99$ & $1.14$
            & $1.83$ & $35.0$ & $1.14$  & $50.9$          \cr
      5     & $5/3$  & $5.0$  & $0.5$ & $1.1$ & $-$
            & $-$    & $-$    & $1.0$   & $50.9$          \cr
      6     & $4/3$  & $5.0$  & $0.5$ & $0.99$ & $1.14$
            & $-$    & $-$    & $1.0$ & $50.9$          \cr
      7     & $2$    & $5.0$  & $0.5$ & $0.99$ & $1.14$
            & $-$    & $-$    & $1.0$ & $50.9$          \cr
\hline
\multicolumn{10}{c}{}\\
\multicolumn{10}{c}{}\\
\end{tabular}

\end{center}

{\bf Table 1.-} $\gamma$ is the adiabatic exponent of the fluid,
 ${\cal M}_{\infty}$ is the asymptotic Mach number,
 $v_{\infty}$ is the asymptotic flow velocity,
 $a$ is the Kerr angular momentum parameter,
 $r_{+}$ indicates the location of the horizon and
 $r_{min}$ and $r_{max}$ are the
 minimum and maximum radial values of the computational domain.
 BL stands for Boyer-Lindquist and KS for Kerr-Schild.
 All distances are measured in units of the mass of the hole,
 which we chose equal to 1. Note that model 5 represents a
 ``naked" singularity $(a>1)$.

\vspace{2cm}

\begin{center}
    \begin{tabular}{|ccccc|}
    \multicolumn{5}{c}{\bf Table 2 \rm} \\
    \multicolumn{5}{c}{\bf Accretion rates versus fluid adiabatic index \rm} \\
    \multicolumn{5}{c}{}\\
    \hline
      MODEL &
      $\gamma$ &
      mass &
      linear momentum &
      angular momentum \cr
\hline
\hline
      4     & $4/3$  & $0.94$  & $-0.98$ & $-0.91$  \cr 
      4     & $5/3$  & $0.83$  & $-0.98$ & $-0.69$ \cr
      4     & $2$    & $0.74$  & $-0.85$ & $-0.67$ \cr
\hline
\multicolumn{5}{c}{}\\
\multicolumn{5}{c}{}\\
\end{tabular}

\end{center}

{\bf Table 2.-} Dependence of the different accretion rates on the
adiabatic index of the fluid for model 4 ($a=0.99M$). The results
are for KS coordinates.  All rates are averaged on the final $200M$ 
of the simulation and conveniently scaled to one.

\vfill

\newpage

%
% Figures
%

\begin{figure}
\centerline{\psfig{figure=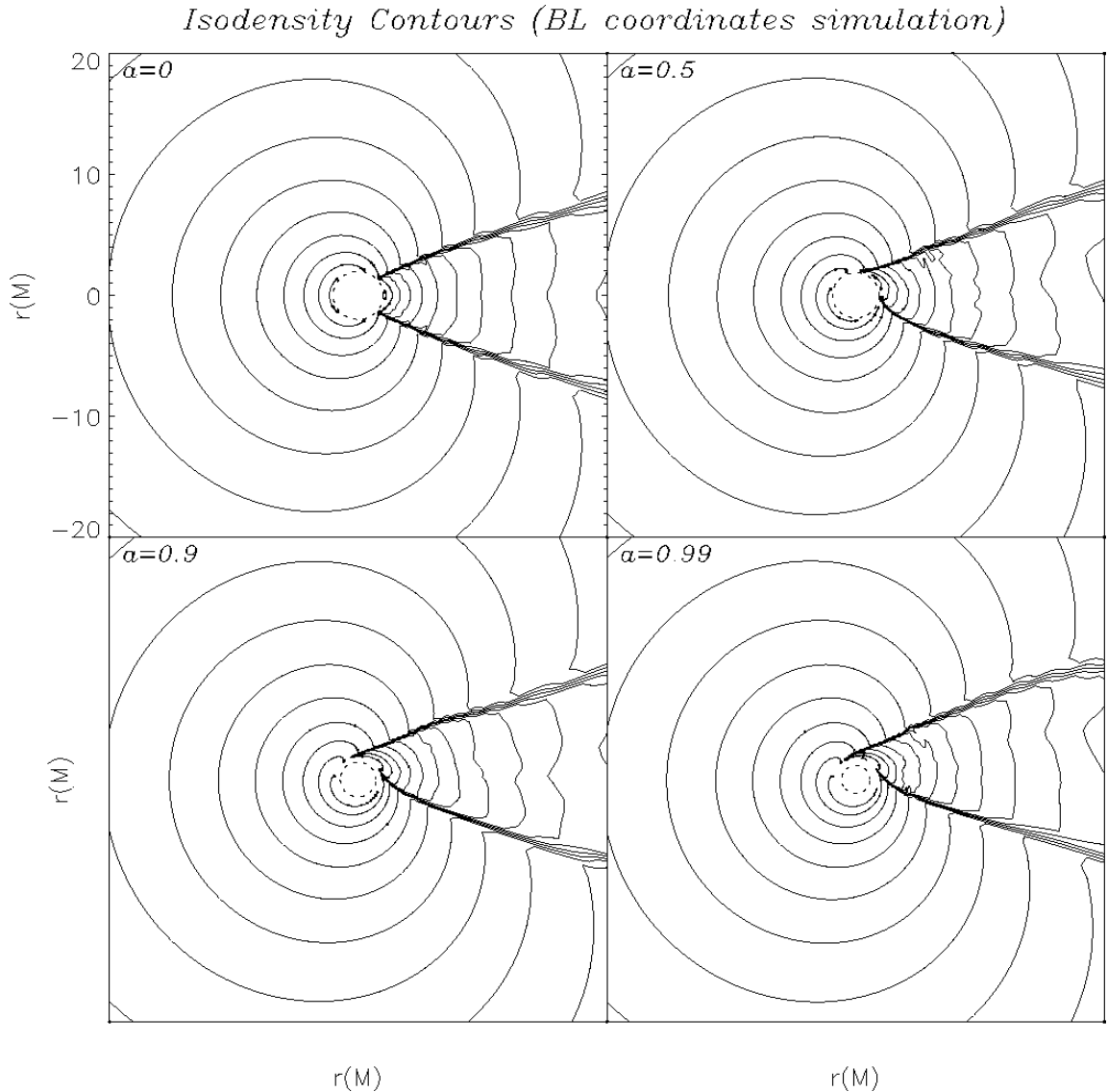,width=7.0in,height=7.0in}}
\caption{{ \protect \small Flow morphology at a final time $t=500M$. We
plot 20 isocontours of the logarithm of the density (scaled to its
asymptotic value) for the first four
models considered in Table 1 ($a=0,0.5,0.9$ and $0.99$, in units of
the mass of the black hole). The presence of a well-defined tail shock
is clearly noticeable in all cases.
The minimum value of $\log\rho$ for all models is $-0.26$
whereas the maximum, always found in the rear (right) part of the black hole,
increases with $a$: $1.65$, $1.69$, $1.75$ and $1.77$ for models 1
to 4, respectively.  The domain of the plot extends up to $20M$. 
Asymptotically the flow moves from left to right. By the final
time of $500M$ the flow pattern is already stationary. 
This figure corresponds to a simulation performed with Boyer-Lindquist 
coordinates. The innermost dashed line indicates the position of the
black hole horizon.
}}
\label{fig1}
\end{figure}

\newpage

\begin{figure}
\centerline{\psfig{figure=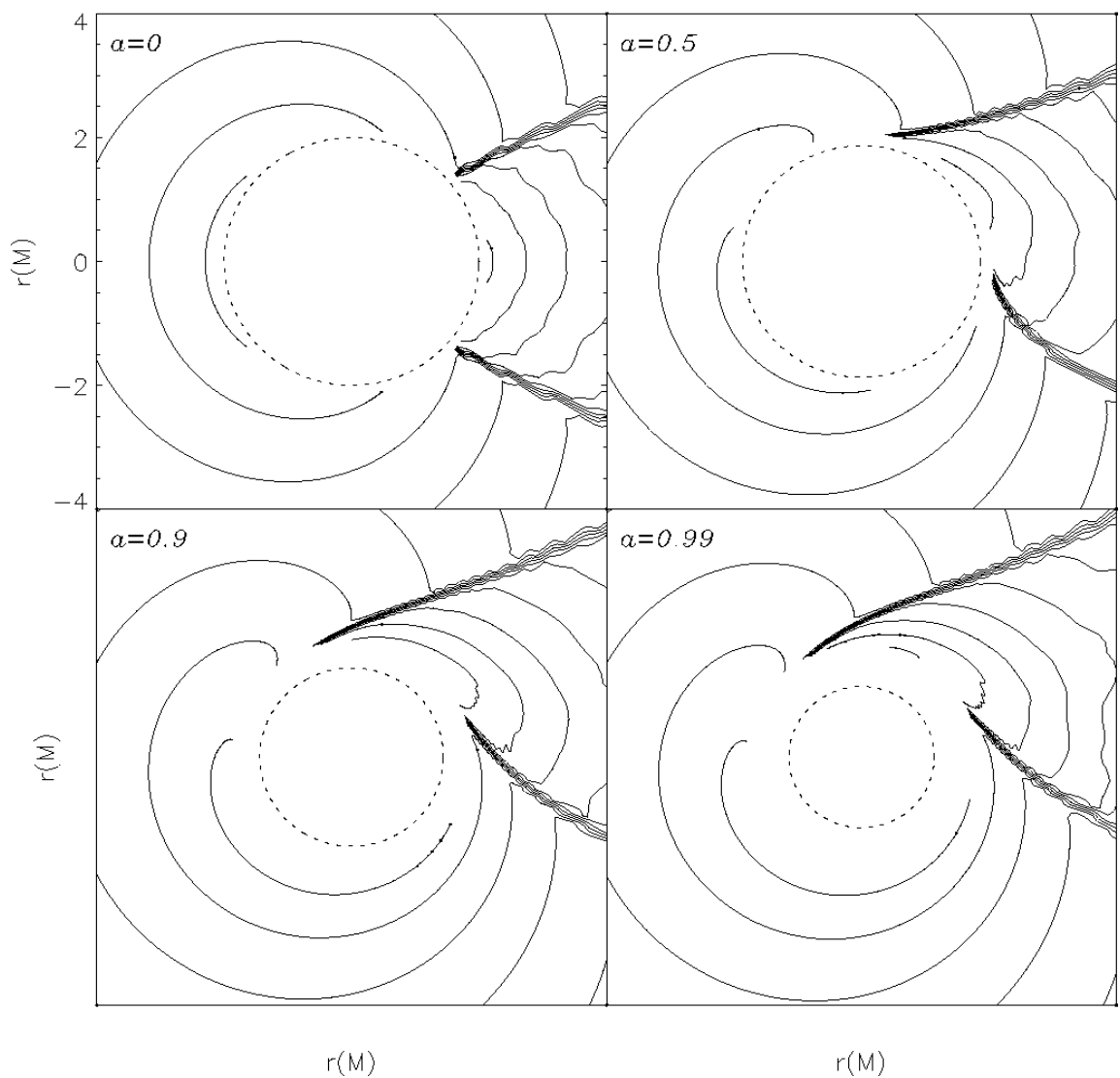,width=7.0in,height=7.0in}}
\caption{{ \protect \small 
This figure represents a zoomed view of Fig.~1. Now the domain of
the plot extends up to $4M$. The effects of the rotation of the
black hole on the flow morphology become increasingly evident
in the regions closer to the black hole horizon. Its location is indicated
by the dashed line. Note that the inner radius of the computational domain
is, in BL coordinates, clearly separated from $r_+$, specially for
rapidly-rotating holes.
}}
\label{fig2}
\end{figure}

\newpage

\begin{figure}
\centerline{\psfig{figure=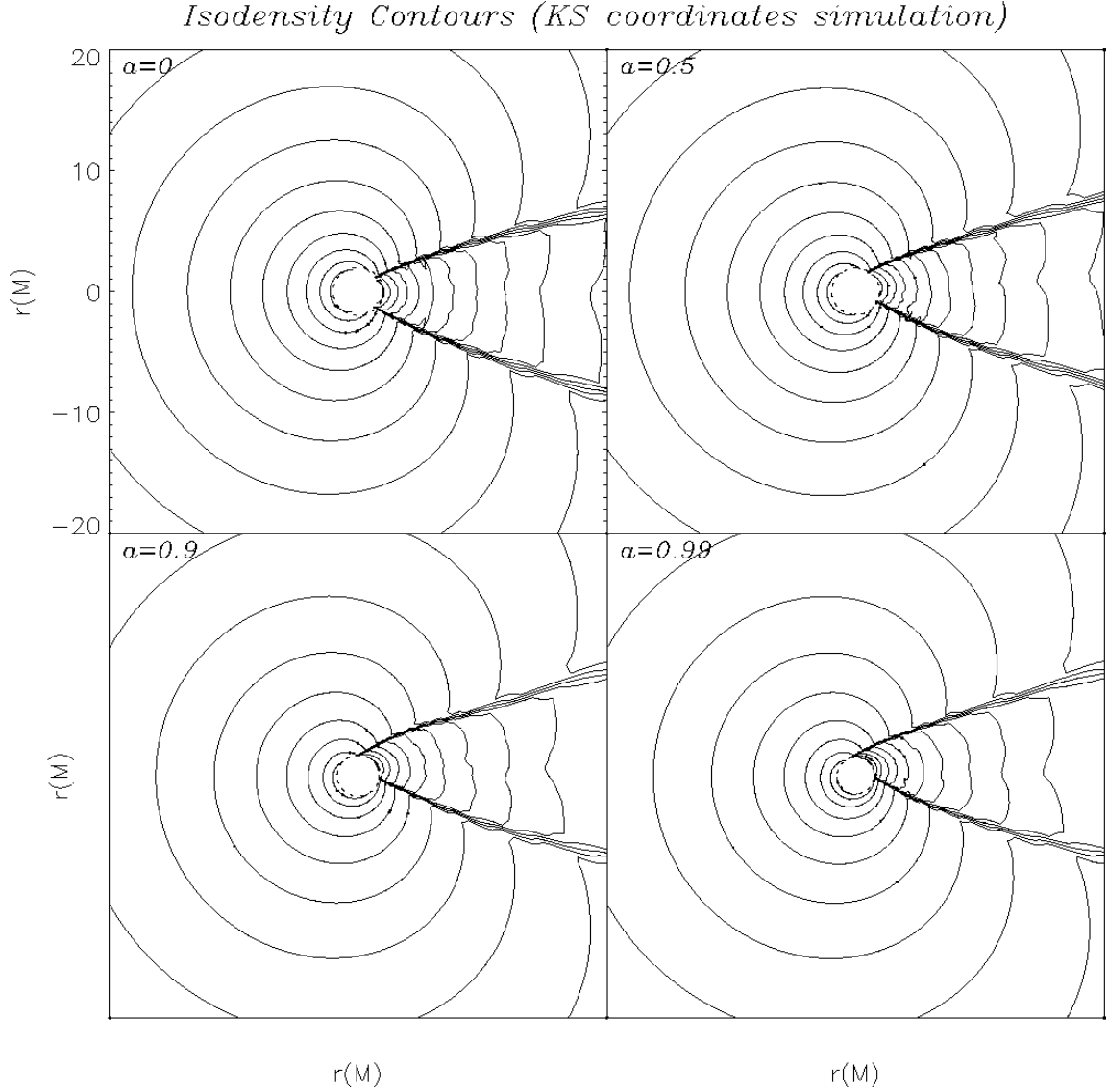,width=7.0in,height=7.0in}}
\caption{{ \protect \small 
Flow morphology at $t=500M$ for a simulation using Kerr-Schild coordinates.
As in Fig.~1 we plot 20 isocontours of the logarithm of the scaled
rest-mass density for the first four models of Table 1. The minimum value of
$\log\rho$ for all models is now $-0.16$. The maximum, again, increases
with $a$: $1.96$, $1.97$, $2.13$ and $2.27$ for models 1 to 4,
respectively. The domain of the plot extends up to $20M$.
The horizon of the black hole is now included in the computational domain.
The shock extends all the way to the horizon, $r_+$,
which is indicated by the innermost dashed line.
}}
\label{fig3}
\end{figure}

\newpage

\begin{figure}
\centerline{\psfig{figure=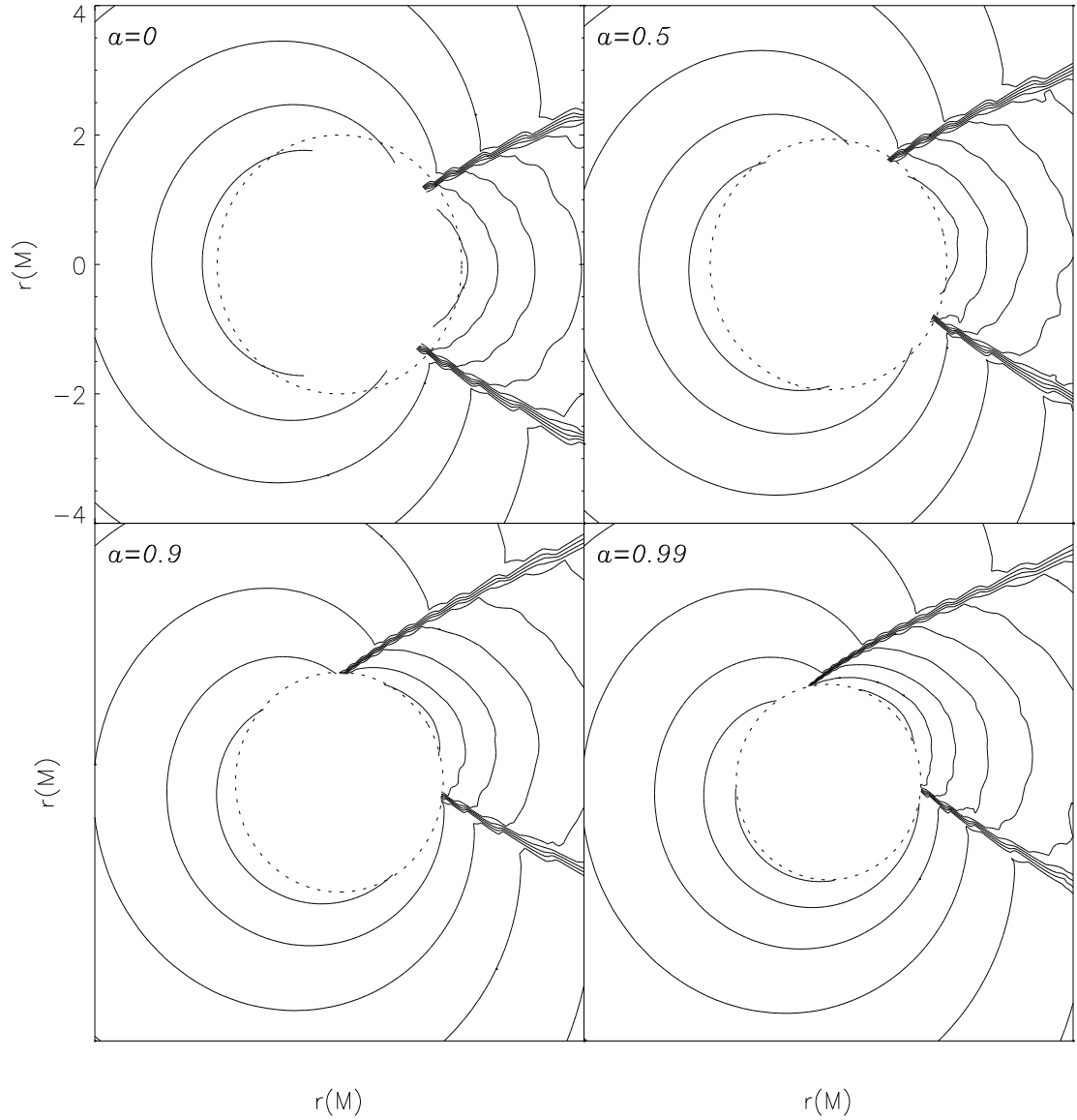,width=7.0in,height=7.0in}}
\caption{{ \protect \small 
This figure represents a zoomed view of Fig.~3. Now the domain of
the plot extends up to $4M$. As in Fig.~2, the effects of the
angular momentum of the black hole on the accreting matter are more
noticeable at the black hole close vicinity. The horizon is again represented
by the innermost dashed line. Contrary to the BL evolutions, the
inner radius of the domain includes now the black hole horizon. Notice
that the presence of the horizon is totally ``transparent" to the 
matter flow.
}}
\label{fig4}
\end{figure}

\newpage

\begin{figure}
\centerline{\psfig{figure=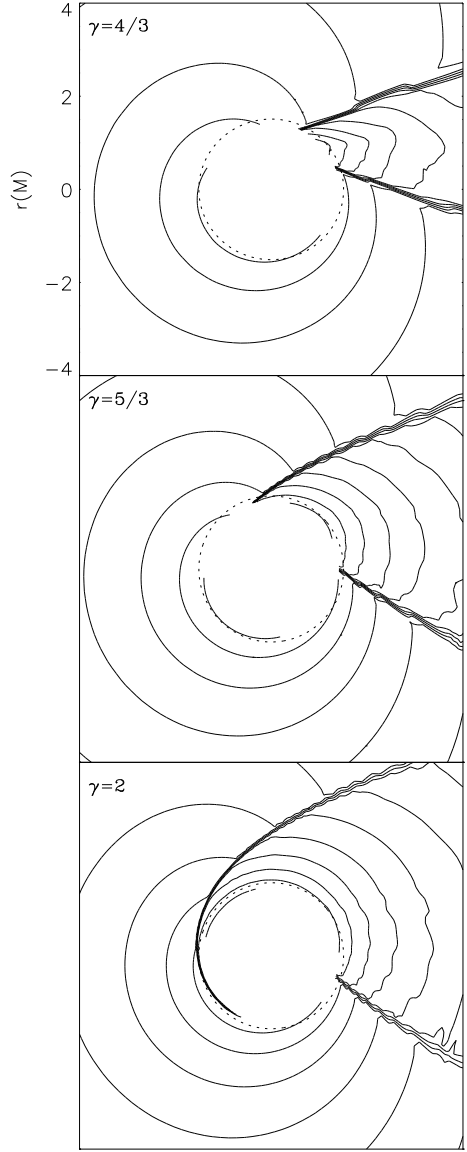,width=2.5in,height=7.5in}}
\caption{{ \protect \small
Stationary flow morphology for three different values of $\gamma$,
the fluid adiabatic index: $4/3$ (top), $5/3$ (middle) and $2$ (bottom).
The solution is depicted at a final time $t=500M$. In every case we plot
20 isocontours of the logarithm of the normalized density from
$-0.18$ to $2.73$ ($\gamma=4/3$), $-0.16$ to $2.27$ ($\gamma=5/3$) and
$-0.15$ to $2.22$ ($\gamma=2$).
The domain extends from $-4M$ to $4M$. The black hole horizon is indicated
by the dashed circle. The larger the value of
$\gamma$ the larger the shock opening angle is and the more wrapped is
the upper shock around the black hole.
}}
\label{3gammas}
\end{figure}

\newpage

\begin{figure}
\centerline{\psfig{figure=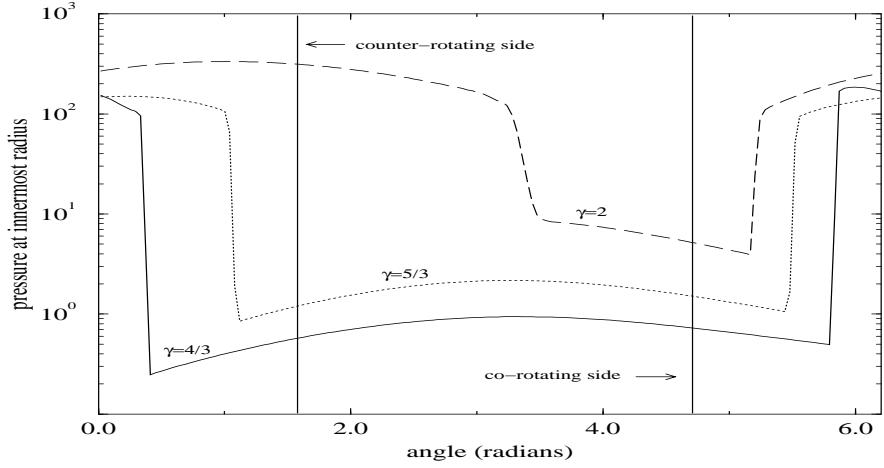,width=5.5in,height=2.8in}}
\caption{{ \protect \small
Pressure profile at the innermost radius for the $a=0.99$ case. Three
different values of $\gamma$ are depicted: $\gamma=4/3$ (solid line),
$\gamma=5/3$ (dotted line) and $\gamma=2$ (dashed line). Note the pressure
difference between the counter and co-rotating sides for the $\gamma=2$
case. This originates a lift force on the black hole. This mechanism is 
analogous to the Magnus effect of classical fluid mechanics.
}}
\label{magnus}
\end{figure}

\begin{figure}
\centerline{\psfig{figure=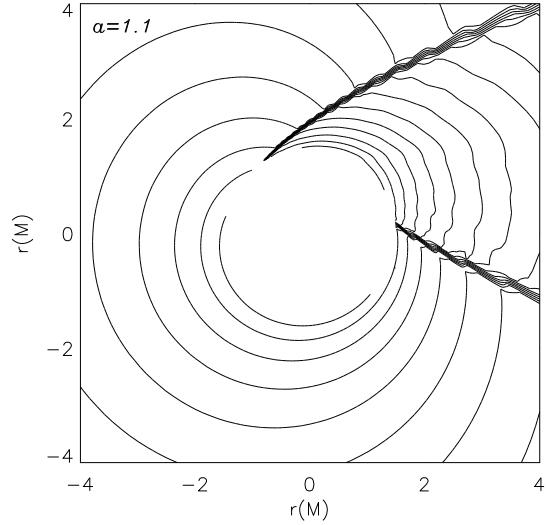,width=4.0in,height=4.0in}}
\caption{{ \protect \small
Morphology of the supersonic accretion flow onto a {\it naked singularity}
($a=1.1M$, model 5) at a final time $t=500M$. Shown are 20 isocontours of the 
logarithm of the normalized density from $-0.16$ to $2.39$. The
domain extends from $-4M$ to $4M$. Note that the flow solution follows the
same trend of Fig.~4: the upper part of the shock wave moves a bit more towards 
the front of the ``hole" as a consequence of both, the larger spin and
pressure gradient at the rear part.
}}
\label{a11}
\end{figure}

\newpage

\begin{figure}
\centerline{\psfig{figure=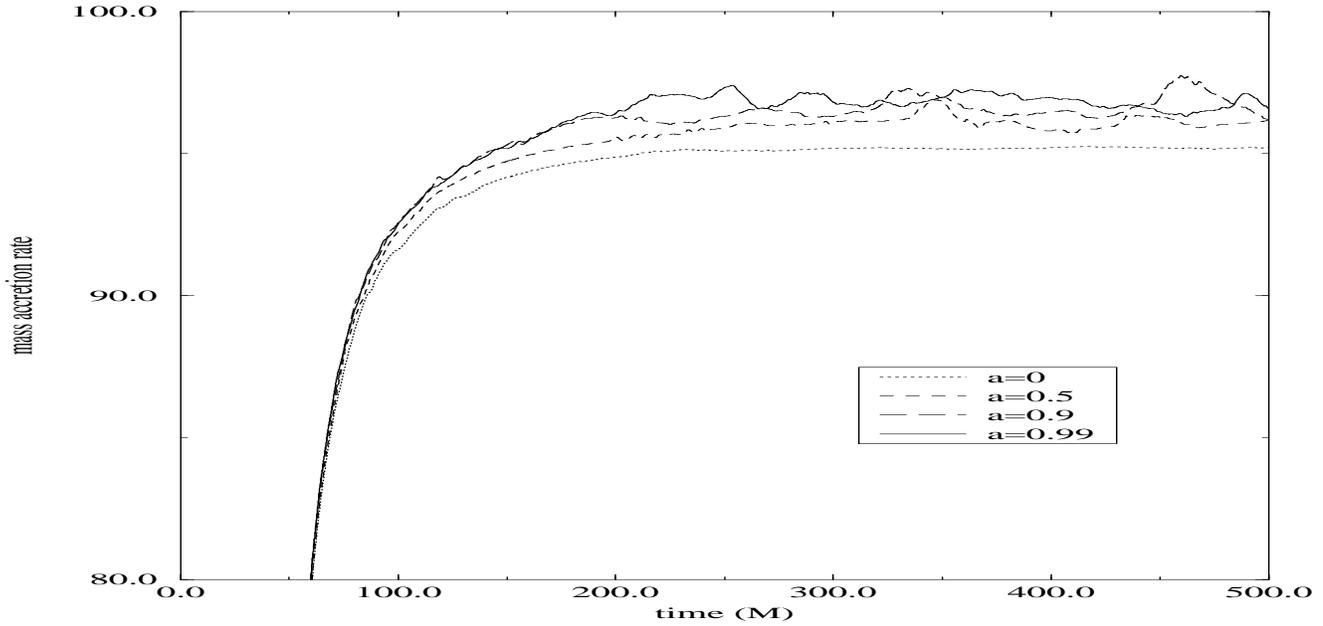,width=8.0in,height=3.4in}}
\caption{{ \protect \small Normalized mass accretion rates as a function of
(coordinate) time for the first four models in Table 1. The figure 
corresponds to the simulation employing Boyer-Lindquist coordinates.
Regardless of the angular momentum of the black hole all models present,
as expected, a similar final steady-state rate. By averaging over the final
$200M$ time interval the maximum differences are found to be of $1.5\%$.
For clarity purposes the $y$ axis has been offset. The non-rotating model
is remarkably stable. The amplitude of the oscillations increase as $a$ 
increases.
}}
\label{fig5}
\end{figure}

%\newpage

\begin{figure}
\centerline{\psfig{figure=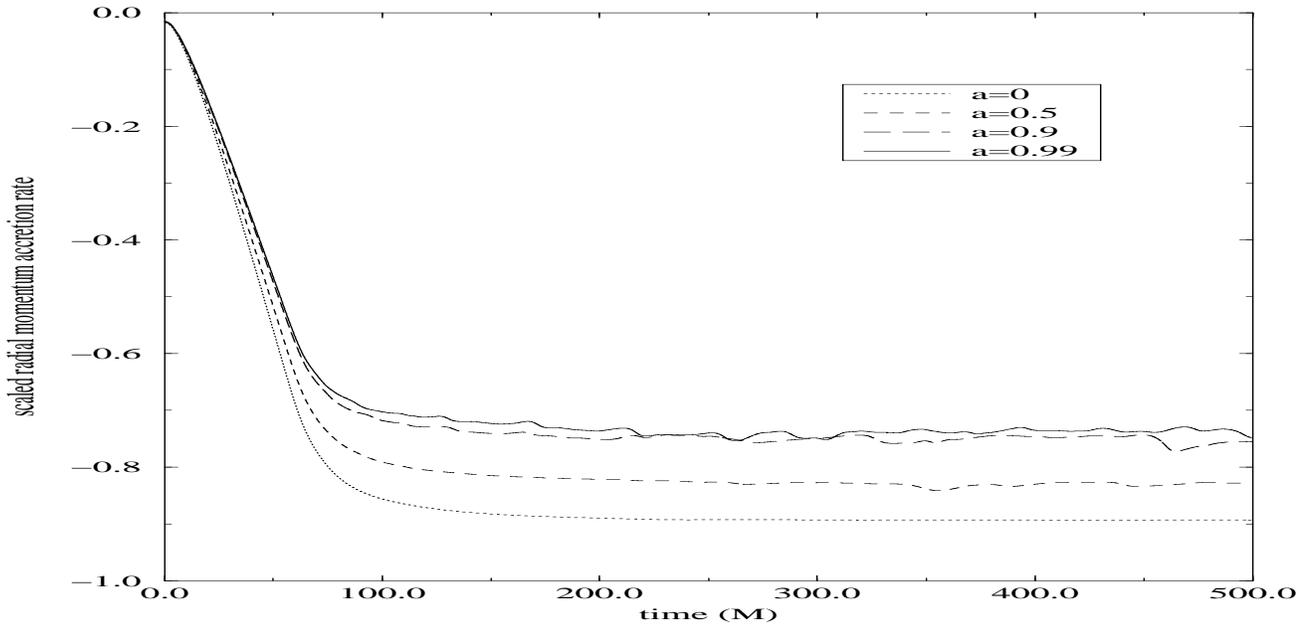,width=8.0in,height=3.4in}}
\caption{{ \protect \small Scaled radial
momentum accretion rates as a function of (coordinate) time for the first 
four models of Table 1. The figure corresponds to the simulation employing 
Boyer-Lindquist coordinates.  As already shown in the mass accretion rate 
(Fig.~8) all models present a similar final steady-state rate. Again, the
non-rotating model shows the most stable behaviour.
}}
\label{fig6}
\end{figure}

\newpage

\begin{figure}
\centerline{\psfig{figure=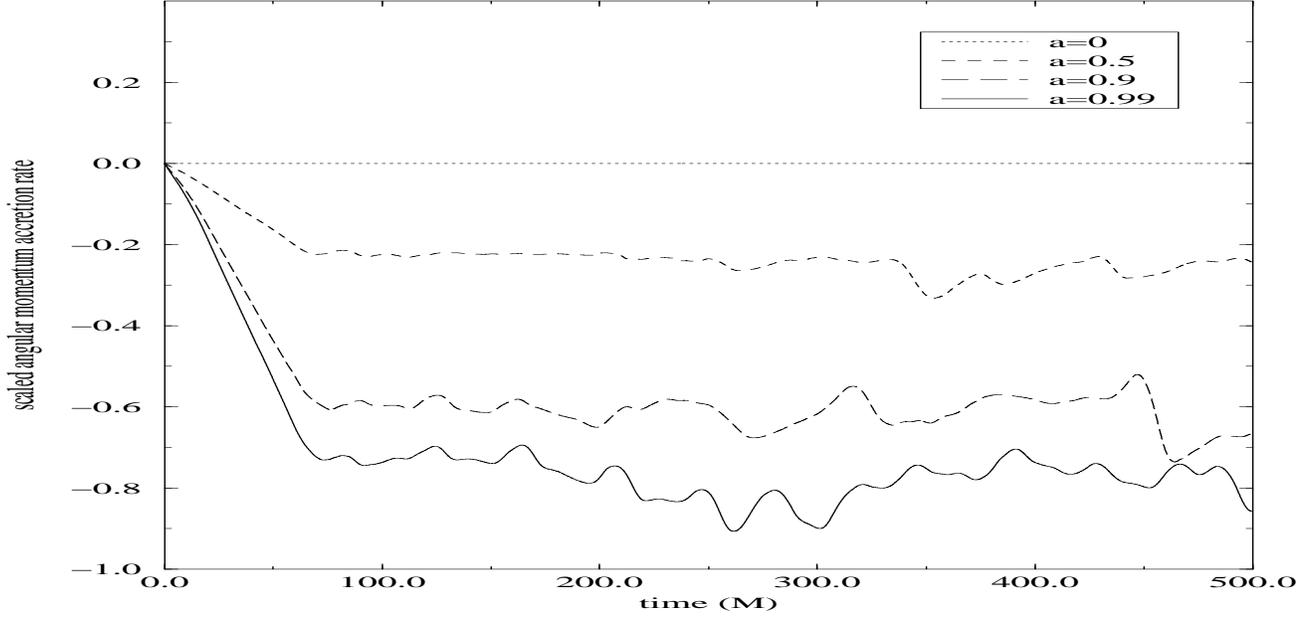,width=8.0in,height=3.4in}}
\caption{{ \protect \small Scaled angular
momentum accretion rates as a function of
(coordinate) time for the first four models of Table 1. 
The figure corresponds to the simulation employing Boyer-Lindquist coordinates.
The accretion rate of angular momentum also proceeds in a stationary
way. It vanishes for Schwarzshild black holes and
increases as the angular momentum of the black hole increases.
}}
\label{fig7}
\end{figure}

%\newpage

\begin{figure}
\centerline{\psfig{figure=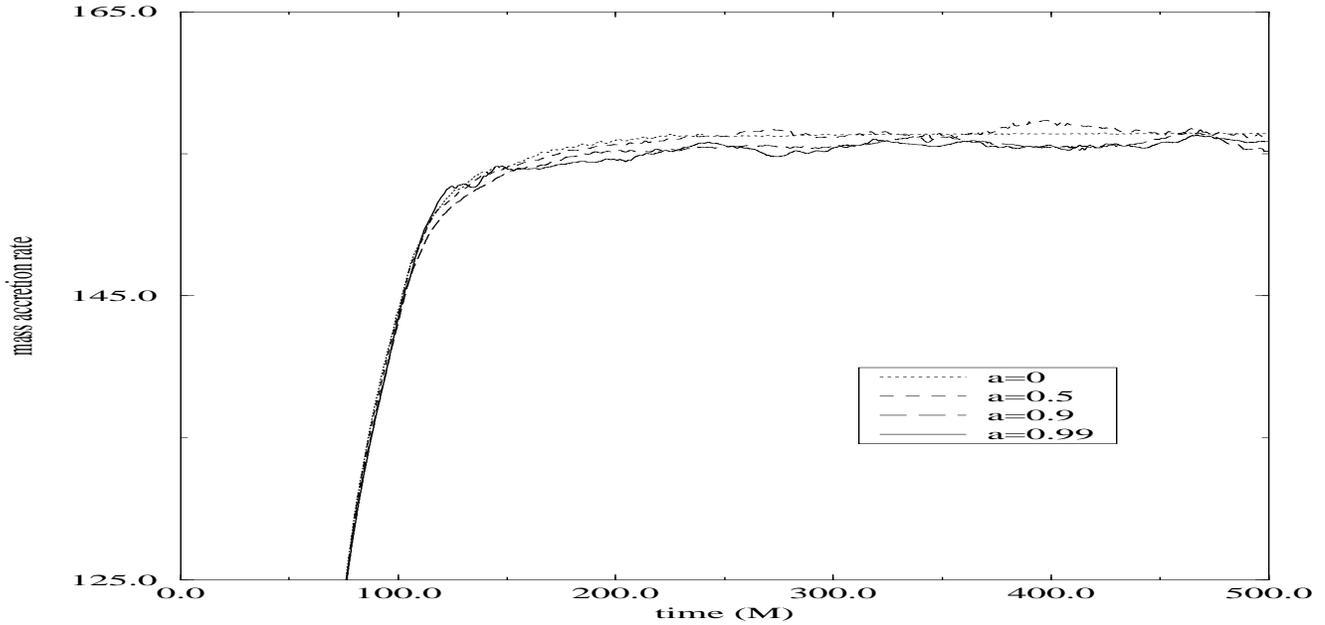,width=8.0in,height=3.4in}}
\caption{{ \protect \small 
Normalized mass accretion rates as a function of (coordinate)
time for the first four models of Table 1. The figure corresponds to the
simulation employing Kerr-Schild coordinates.  For clarity purposes the $y$ axis 
has been offset. All models present a fairly constant value, being the maximum
differences less than $1.2\%$.
}}
\label{fig8}
\end{figure}

\newpage

\begin{figure}
\centerline{\psfig{figure=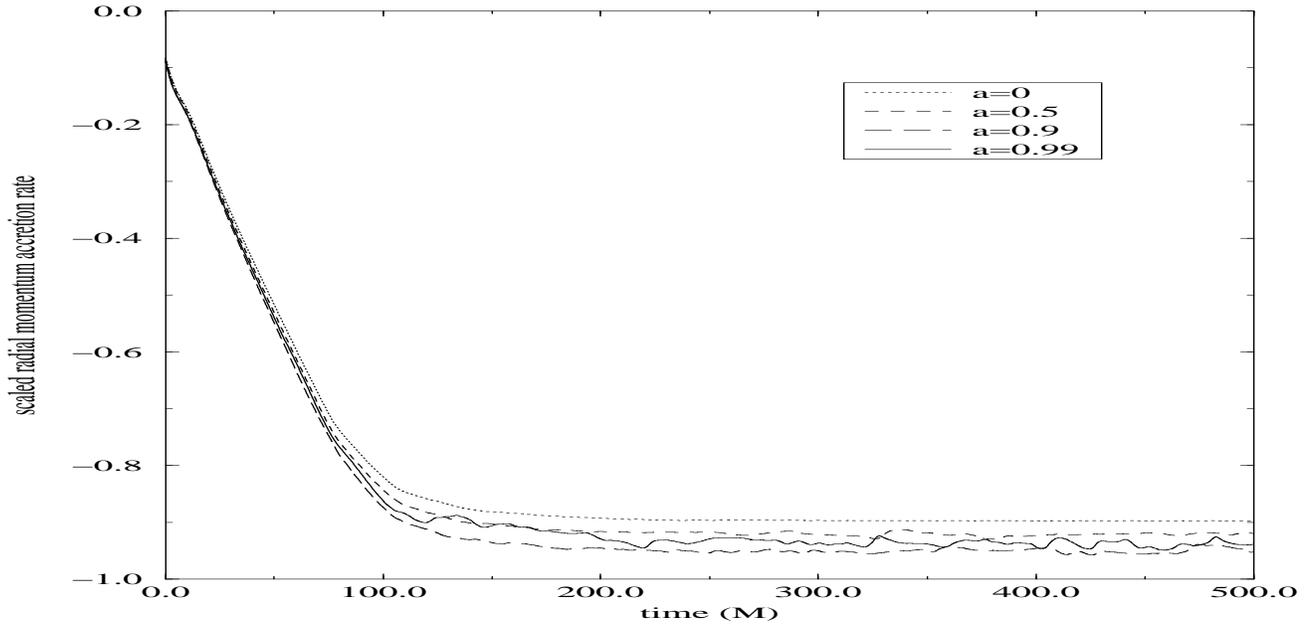,width=8.0in,height=3.4in}}
\caption{{ \protect \small 
Scaled radial momentum accretion rates as a function of (coordinate)
time for the first four models of Table 1. The figure corresponds to the
simulation employing Kerr-Schild coordinates. Again, it is noticeable the
stability achieved for the non-rotating case.
}}
\label{fig9}
\end{figure}

%\newpage

\begin{figure}
\centerline{\psfig{figure=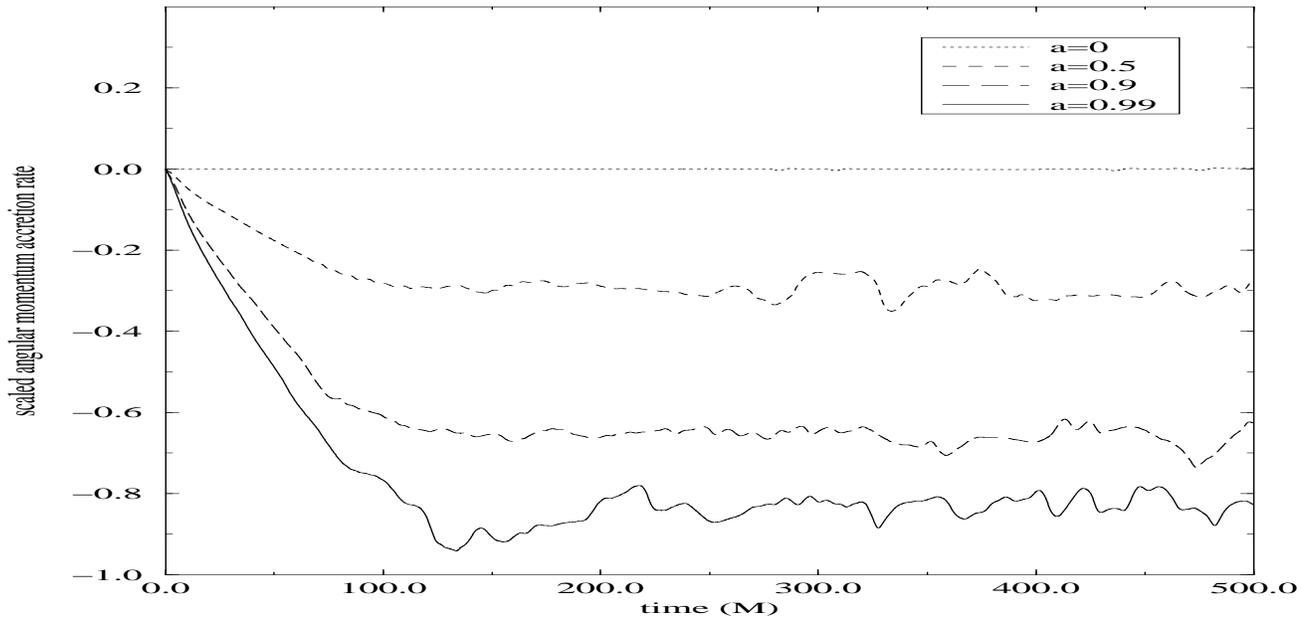,width=8.0in,height=3.4in}}
\caption{{ \protect \small 
Scaled angular momentum accretion rates as a function of (coordinate) time 
for the first four models of Table 1. The figure corresponds to
the simulation employing Kerr-Schild coordinates.  
}}
\label{fig10}
\end{figure}

\newpage
\begin{figure}
\centerline{\psfig{figure=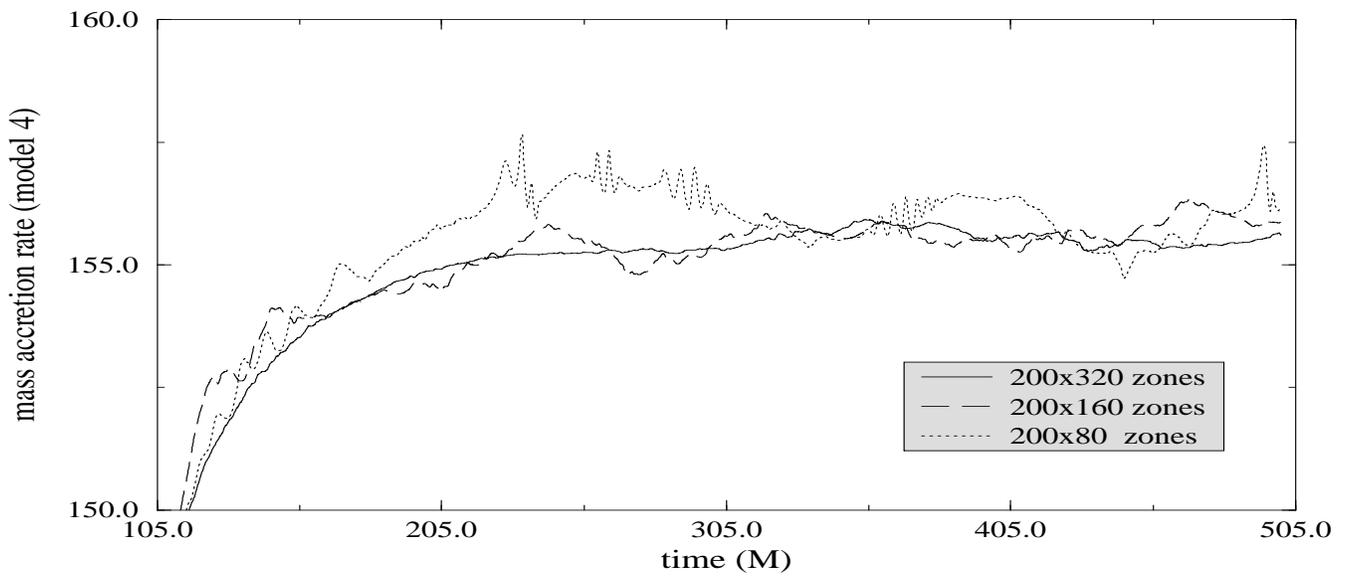,width=8.0in,height=3.4in}}
\caption{{ \protect \small
Mass accretion rate as a function of (coordinate) time for model 4 of
Table 1. All models with $a\ne0$ present a much more oscillatory 
behaviour than the non-rotating one.  This figure shows results obtained 
with three different angular resolutions. As the resolution is increased
the amplitude of the oscillations is significantly reduced, which is
a clear indication of the numerical nature of such oscillations. Note that,
for clarity, both axis have been offset.
}}
\label{figconv}
\end{figure}

\newpage

\begin{figure}
\centerline{\psfig{figure=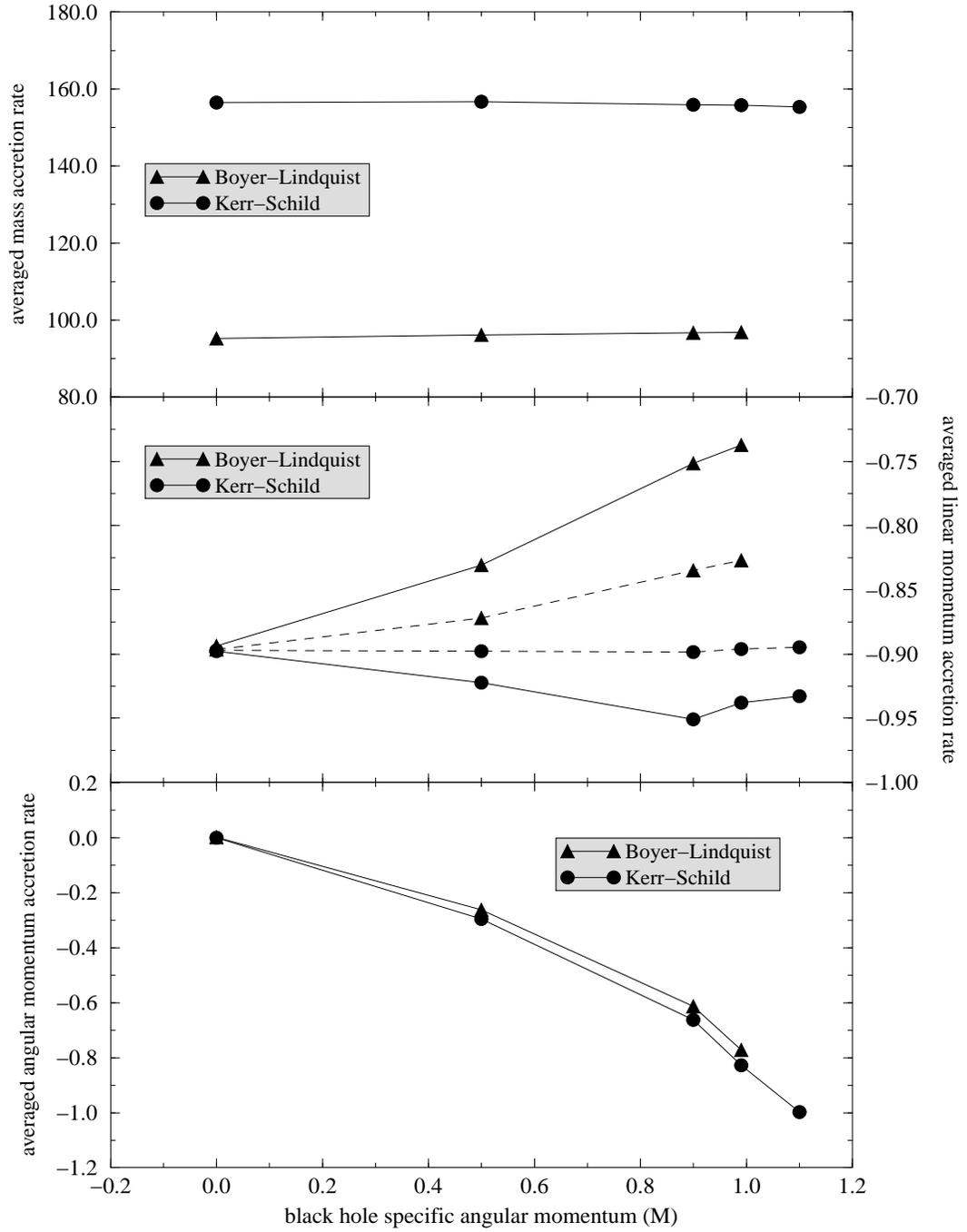,width=5.0in,height=7.5in}}
\caption{{ \protect \small Averaged accretion rates of mass (top), linear
momentum (middle) and angular momentum (bottom) versus the black hole spin.
All rates are averaged at the final $200M$ of the evolution. They are all
computed at the accretion radius except for the linear momentum rate which
is also computed at $r_{max}$ (dashed lines). The non-dependence of the
mass and linear momentum rates with $a$ is noticeable, especially for the
KS coordinate system. For the linear momentum rate this is particularly
true when computed at large radii ($r_{max}$, dashed lines)
but does not hold, for numerical reasons, at small radii ($r_a$, solid lines).
}}
\label{average}
\end{figure}

\newpage

\begin{figure}
\centerline{\psfig{figure=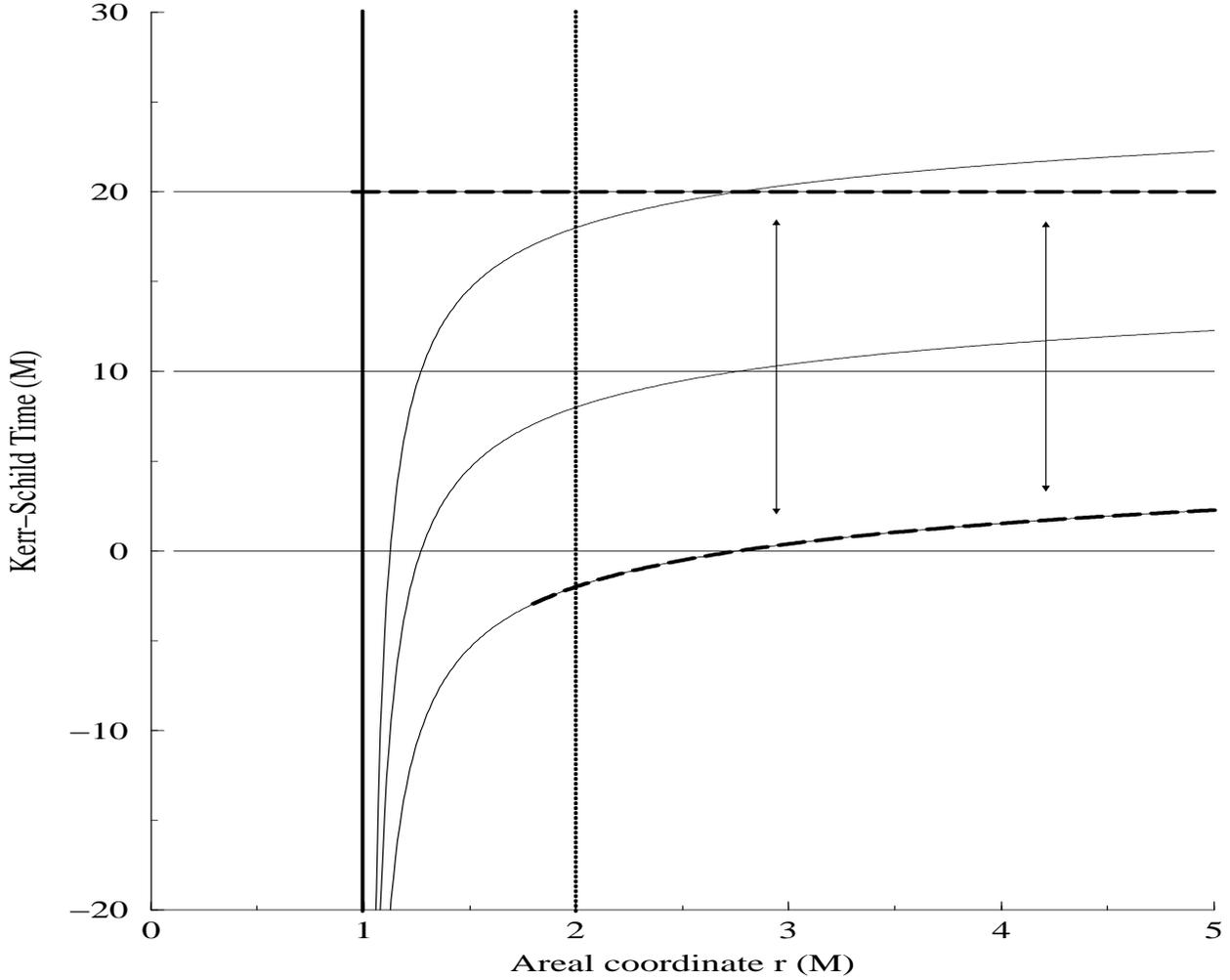,width=8.0in,height=5.5in}}
\caption{{ \protect \small Schematic illustration of the setup that
allows for a simple comparison between the Boyer-Lindquist (BL) and
Kerr-Schild (KS) coordinate systems in the case of stationary
flows. The black hole is assumed maximally rotating, the hypersurfaces
depicted lie in the equatorial plane and the azimuthal direction is
suppressed.  Shown are three different KS time levels at intervals of
10M (thin horizontal lines) and the corresponding BL levels (curved
lines). The BL time levels retard infinitely long at the horizon
(thick vertical line at 1M). The thick long dashed lines represent the
domains on which flows are being computed in both coordinate systems.
The KS flow extends just inside the horizon, whereas the BL flow is
truncated at about $1.8M$ (the dotted vertical line at 2M is the
ergosphere boundary). For stationary flows, a one-to-one
correspondence between physical points can be established using the
appropriate coordinate transformations.  }}
\label{mapping}
\end{figure}

\newpage

\begin{figure}
\centerline{\psfig{figure=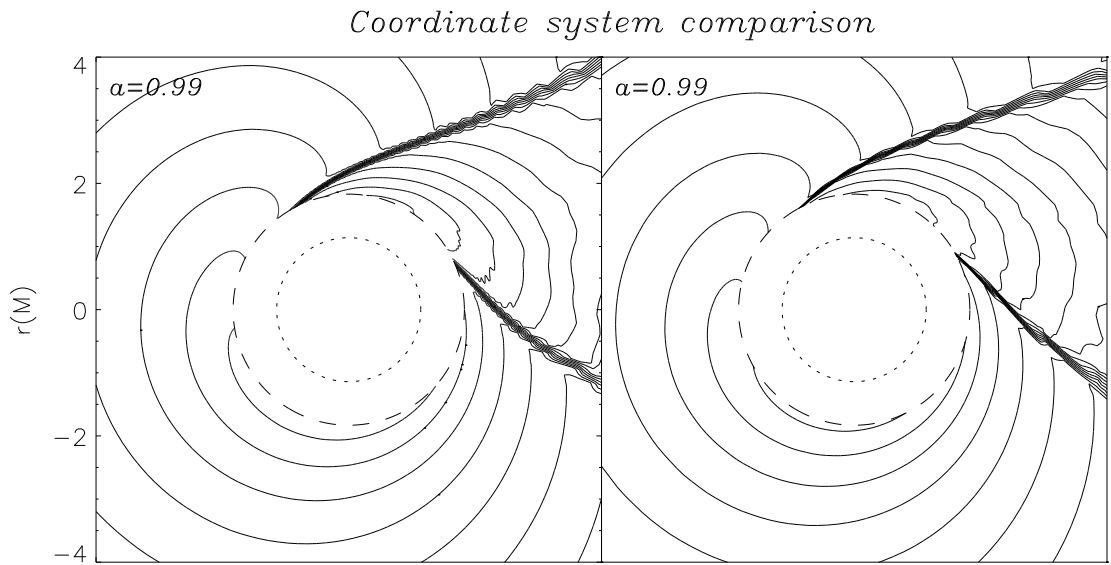,width=7.0in,height=3.5in}}
\caption{{ \protect \small Comparison of the stationary accretion pattern.
Isocontours of the logarithm of the density for Model 4 ($a=0.99$).
Left panel shows the Boyer-Lindquist evolution. The right panel
corresponds to a Kerr-Schild integration but transformed back to
Boyer-Lindquist coordinates. The dotted and dashed circles indicate the
location of the horizon and $r_{min}$, respectively. The agreement
is remarkable.
}}
\label{fig11}
\end{figure}

\newpage

\begin{figure}
\centerline{\psfig{figure=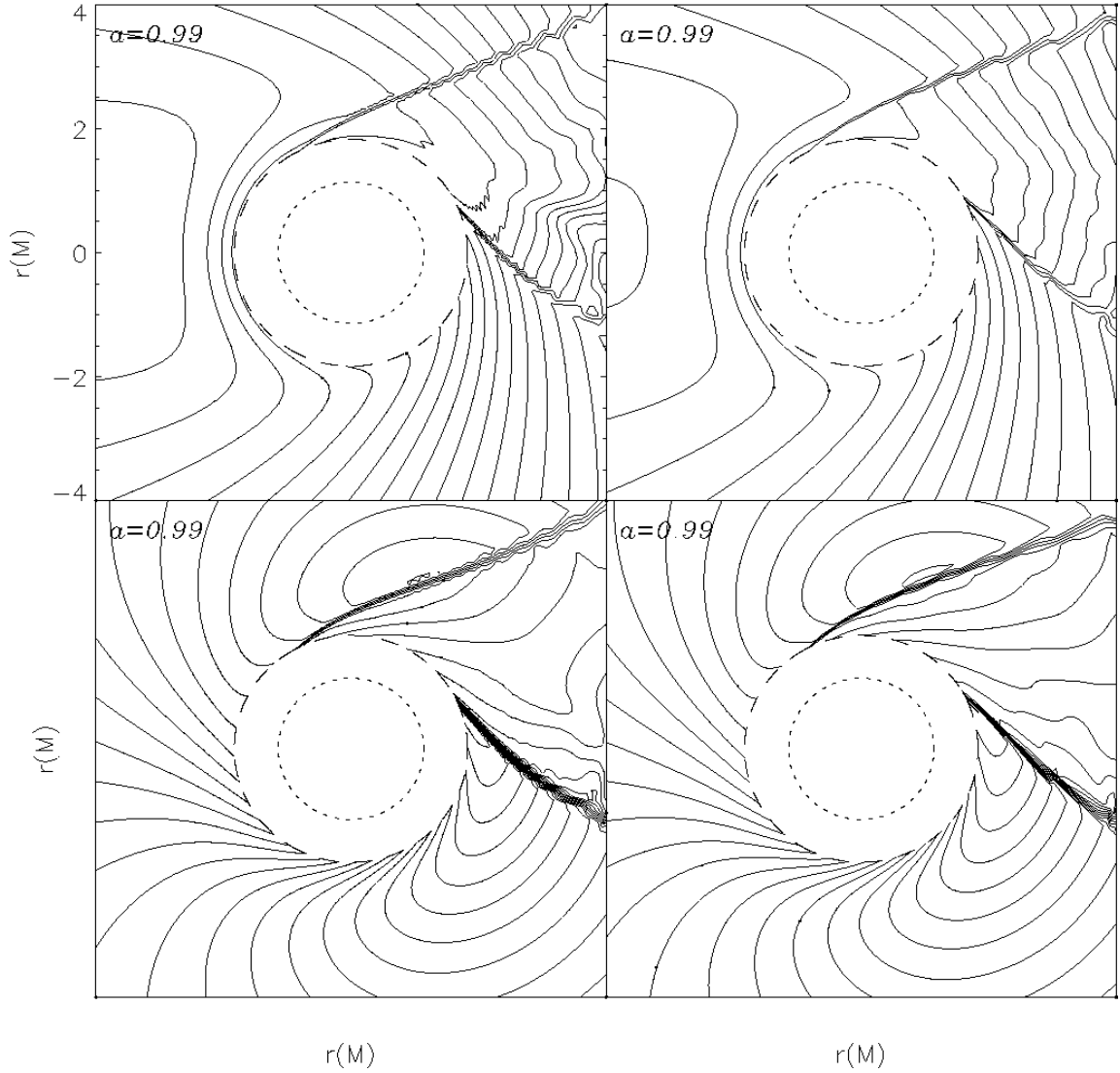,width=7.0in,height=7.0in}}
\caption{{ \protect \small Comparison of the stationary accretion pattern
(model 4). Isocontours of the radial (top) and azimuthal (bottom) 3-velocity
components. The left and right panels follow the same convention
of Fig.~17. The agreement in the simulations is again excellent.
}}
\label{fig12}
\end{figure}

\clearpage

\newpage

\begin{figure}
\centerline{\psfig{figure=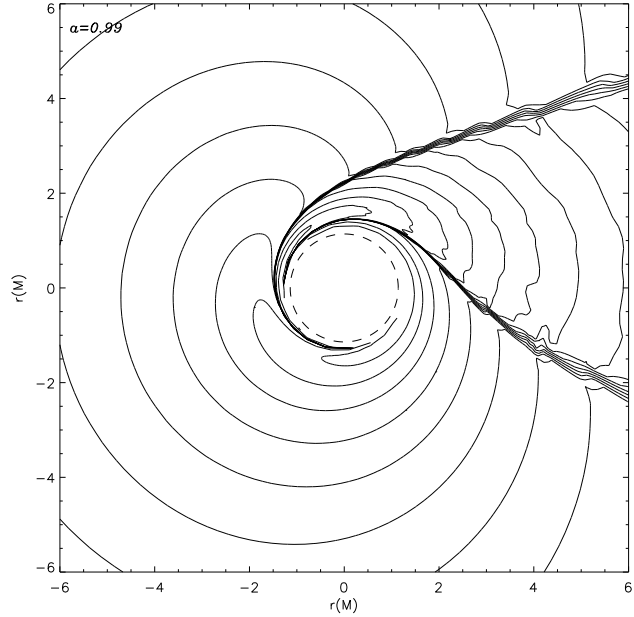,width=3.5in,height=3.5in}}
\caption{{ \protect \small Extension of the stationary accretion
pattern of Fig.~17 (right) to regions much closer to the 
black hole horizon. In Boyer-Lindquist coordinates the shock is totally 
wrapped around the horizon.
}}
\label{fig13}
\end{figure}

\begin{figure}
\centerline{\psfig{figure=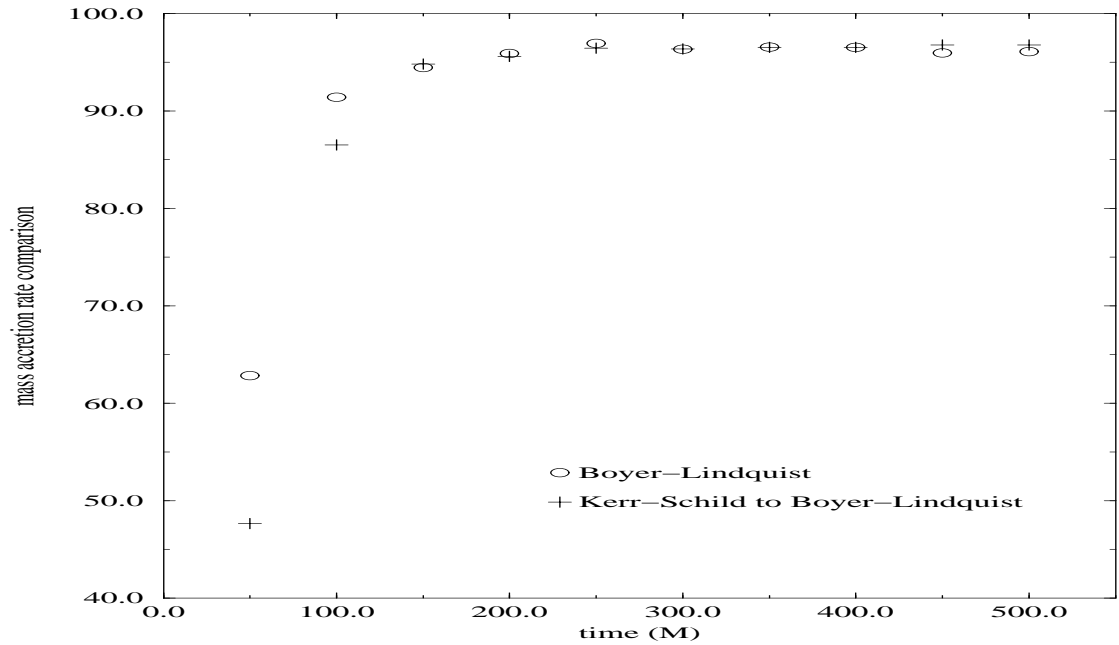,width=7.0in,height=3.5in}}
\caption{{ \protect \small Comparison of the mass accretion rates for
model 4 as computed originally in Boyer-Lindquist coordinates (circles) or 
transformed from Kerr-Schild coordinates (plus signs). This plot clearly 
demonstrates and quantifies the large amount of agreement found in the 
simulations.
}}
\label{fig16}
\end{figure}

\end{document}